\documentclass{emulateapj}

\usepackage{ccaption}

\newcommand\kl{{\rm{k}\lambda}}
\newcommand\ea{et al.\ }
\newcommand\ah{^{\rm h}}
\newcommand\am{^{\rm m}}
\newcommand\as{^{\rm s}}
\newcommand\pr{^{\prime}}
\newcommand\twopr{^{\prime \prime}}
\slugcomment{Submitted to {\em Astrophysical Journal}}

\shorttitle{CMB Anisotropy}
\shortauthors{Dawson \ea}

\begin{document}

\title{Final Results from the BIMA CMB Anisotropy Survey and Search for Signature of the SZ effect}

\author{
K.S.~Dawson\altaffilmark{1},
W.L.~Holzapfel\altaffilmark{1},
J.E.~Carlstrom\altaffilmark{2,3},\\
M.~Joy\altaffilmark{4},
and S.J.~LaRoque\altaffilmark{2}}

%\altaffiltext{1}{E.O. Lawrence Berkeley National Laboratory, 1 Cyclotron Rd.,
%Berkeley CA 94720}
\altaffiltext{1}{Department of Physics, University of California,
Berkeley CA 94720 (KD now at LBNL)}
\altaffiltext{2}{Kavli Institute for Cosmological Physics, Department of Astronomy and Astrophysics, University of Chicago, Chicago
IL 60637}
\altaffiltext{3}{Department of Physics, Enrico Fermi Institute,
University of Chicago, Chicago, IL 60637}
\altaffiltext{4}{Department of Space Science, XD12, NASA Marshall Space Flight Center,
Huntsville AL 35812}
\email{kdawson@lbl.gov}

\begin{abstract}
We report the final results of our study of the cosmic microwave background (CMB)
with the BIMA array.
Over 1000 hours of observation were dedicated to this
project exploring CMB anisotropy on scales between $1\pr$ and $2\pr$
in eighteen $6\pr.6$ FWHM fields.  
In the analysis of the CMB power spectrum, the visibility data is divided
into two bins corresponding to different angular scales.
Modeling the observed excess power as a flat
band of average multipole $\ell_{eff} = 5237$, we find $\Delta
T_1^2=220_{-120}^{+140}\,\mu$K$^2$ at $68\%$ confidence and $\Delta
T_1^2 >0 \,\mu$K$^2$ with $94.7\%$ confidence.  In a second band with
average multipole of $\ell_{eff} = 8748$, we find $\Delta T_2^2$
consistent with zero, and an upper limit $880\,\mu$K$^2$ at $95\%$
confidence.  
An extensive series of tests and supplemental observations with the VLA
provide strong evidence against systematic errors or radio point sources 
being the source of the observed excess power.  
The dominant source of anisotropy on these
scales is expected to arise from the Sunyaev-Zel'dovich (SZ) effect in
a population of distant galaxy clusters.  
If the excess power is due to the SZ effect, we can place constraints
on the normalization of the matter power spectrum
$\sigma_8 = 1.03^{+0.20}_{-0.29}$ at $68\%$ confidence.
The distribution of pixel
fluxes in the BIMA images are found to be consistent with simulated
observations of the expected SZ background and rule out instrumental
noise or radio sources as the source of the observed excess 
power with similar confidence to the detection of excess power.  
Follow-up optical
observations to search for galaxy over-densities anti-correlated with
flux in the BIMA images, as might be expected from the SZ effect,
proved to be inconclusive.

\end{abstract}


\section{Introduction}

The angular power spectrum of the Cosmic Microwave Background (CMB)
has been measured with high signal to noise on scales from degrees
to several arcminutes (e.g., Hinshaw \ea 2003, Kuo \ea 2004,
Mason \ea 2003).  However, observations of CMB anisotropy on
arcminute scales, where secondary anisotropies such as the
Sunyaev-Zel'dovich (SZ) effect (Sunyaev \& Zel'dovich 1970) are
expected to dominate the primary CMB anisotropy (e.g., Gnedin \& Jaffe
2001), have not yet achieved comparable precision.  
At these finer angular scales, observations
distant SZ clusters have the potential to
be a powerful probe of the growth of structure in the Universe (Carlstrom, Holder, \& Reese 2002). 
However, the power spectrum of the arcminute fluctuations reveals little information
about the nature of the sources responsible for the anisotropy.  As
suggested by Rubino-Martin \& Sunyaev (2003), higher order statistics
of images can, in principle, be used to identify the unique signature of the SZ
effect.

Beginning in the summer of 1998, we began a program to search for
arcminute-scale CMB anisotropy using the Berkeley-Illinois-Maryland
Association (BIMA\footnote{The BIMA array
is operated with support from the National Science Foundation})
interferometer.  Initial results are included in
earlier publications (Holzapfel \ea 2000, Dawson \ea 2001, and
Dawson \ea 2002 (hereafter D2002)).  A detailed description of the
BIMA analysis, results from other experiments, and a comparison with
theoretical models and simulations of structure formation can be found
in D2002.  In this paper, we report the final results from the BIMA
CMB anisotropy survey.  We describe the observations of the final
fields in summer 2002 with the BIMA array and the Very Large Array
(VLA\footnote{The VLA is operated by the National Radio Astronomy
Observatory, a facility of the National Science Foundation, operated
under cooperative agreement by Associated Universities, Inc.})
in \S\ref{sec:obs}.  The results of the power spectrum analysis,
including a discussion of tests
for systematic errors in the analysis are presented in \S\ref{sec:power}.
In \S\ref{sec:images} we
examine the BIMA image statistics in order to constrain the origin of
the observed anisotropy.  In \S\ref{sec:optical}, we describe the
results of follow-up optical observations used in an attempt to
identify galaxy clusters in the BIMA fields.  Finally, in
\S\ref{sec:con}, we summarize the results and present our conclusions.

\section{Observations}\label{sec:obs}
The BIMA anisotropy survey consists of 18 fields that had not been
previously observed for SZ galaxy clusters or CMB anisotropy at
arcminute angular scales.  Each field was observed with the
BIMA array at a frequency of $28.5$ GHz and the VLA
at a frequency of $4.8$ GHz.  Analysis of ten fields observed with the BIMA
array during the summers of 1998, 2000, and 2001 revealed evidence for
a detection of power in excess of the instrument noise.
A description of the analysis and results for the first ten fields in the 
survey can be found in D2002.  
Eight new fields were added to the survey in the summer of 2002.
 
\subsection{BIMA Observations} \label{subsec:bima}
 
All anisotropy observations were made using the BIMA array at Hat Creek.
Nine $6.1$ meter telescopes of the array
were equipped for operation at $28.5$ GHz, providing a $6.6\pr$ FWHM field of view.
The first ten fields, BDF4-BDF13, were chosen to lie in regions expected to have minimal 
radio point source and dust contamination.
The eight fields added in 2002, BDF14-BDF21, were chosen to lie two minutes east in Right
Ascension of existing fields in the survey.
Observations of these fields added another 489 hours of observations to the previously
published data, for a total of 1096 hours for the complete BIMA survey.
Each of these new fields was observed using the same phase calibrator as its 
previously observed nearest neighbor
and analyzed following the data reduction as described in D2002.
These fields were chosen to check for possible
contamination of the signal correlated with telescope position.
This makes for a total of 18 independent fields
in the survey, covering approximately $0.2$ square degrees.
The pointing center, dates of observation, and observation time for each of the fields
are given in Table \ref{tab:obstimes}.
We dedicated $55-65$ hours of observation with the BIMA array to each of the new fields
in order to achieve a uniform RMS noise level of
$<150\,\mu$Jy/beam on short baselines ($u$-$v < 1.1 \, \kl$) for the entire sample.
This noise level corresponds to an RMS of $15.55 \,\mu$K for a $2\pr$ synthesized beam.

\begin{table*}[!htb]
\caption{\label{tab:obstimes}Field Positions and Observation Times}
\small
\begin{center}
\begin{tabular}{lcccc}
\hline\hline
\multicolumn{1}{c}{Fields} & R. A. (J2000) & Decl. (J2000)  & Observation year(s) & Time (Hrs) \\ \hline
BDF4 & $00\ah\,28\am\,04.4\as$ & $+28^{\circ}\,23\pr\,06\twopr$ & 98 &  $77.6$\\
HDF & $12\ah\,36\am\,49.4\as$ & $+62^{\circ}\,12\pr\,58\twopr$ & 98, 01 &  $59.9$\\
BDF6 & $18\ah\,21\am\,00.0\as$ & $+59^{\circ}\,15\pr\,00\twopr$ & 98, 00 &  $81.2$\\
BDF7 & $06\ah\,58\am\,45.0\as$ & $+55^{\circ}\,17\pr\,00\twopr$ & 98, 00 & $68.2$\\
BDF8 & $00\ah\,17\am\,30.0\as$ & $+29^{\circ}\,00\pr\,00\twopr$ & 00, 01 & $53.3$\\
BDF9 & $12\ah\,50\am\,15.0\as$ & $+56^{\circ}\,52\pr\,30\twopr$ & 00, 01 &  $53.9$\\
BDF10 & $18\ah\,12\am\,37.2\as$ & $+58^{\circ}\,32\pr\,00\twopr$ & 00, 01 & $53.3$\\
BDF11 & $06\ah\,58\am\,00.0\as$ & $+54^{\circ}\,24\pr\,00\twopr$ & 00, 01 & $50.0$\\
BDF12 & $06\ah\,57\am\,38.0\as$ & $+55^{\circ}\,32\pr\,00\twopr$ & 01 &  $54.8$\\
BDF13 & $22\ah\,22\am\,45.0\as$ & $+36^{\circ}\,37\pr\,00\twopr$ & 01 &  $54.5$\\
BDF14 & $00\ah\,26\am\,04.4\as$ & $+28^{\circ}\,23\pr\,06\twopr$ & 02 & $62.4$\\
BDF15 & $06\ah\,56\am\,45.0\as$ & $+55^{\circ}\,17\pr\,00\twopr$ & 02 & $64.2$\\
BDF16 & $12\ah\,34\am\,49.4\as$ & $+62^{\circ}\,12\pr\,58\twopr$ & 02 &  $64.5$\\
BDF17 & $18\ah\,19\am\,00.0\as$ & $+59^{\circ}\,15\pr\,00\twopr$ & 02 &  $64.5$\\
BDF18 & $00\ah\,15\am\,30.0\as$ & $+29^{\circ}\,00\pr\,00\twopr$ & 02 & $57.8$\\
BDF19 & $06\ah\,55\am\,38.0\as$ & $+55^{\circ}\,32\pr\,00\twopr$ & 02 & $59.4$\\
BDF20 & $12\ah\,48\am\,15.0\as$ & $+56^{\circ}\,52\pr\,30\twopr$ & 02 &  $54.7$\\
BDF21 & $18\ah\,10\am\,37.2\as$ & $+58^{\circ}\,32\pr\,00\twopr$ & 02 &  $62.0$\\
\hline
\end{tabular}
\end{center}
\begin{center}
\end{center}
\normalsize
\end{table*}
 
\subsection{VLA Observations} \label{sec:VLA}

To help constrain the contribution from point sources to the anisotropy
measurements, we used the VLA at a frequency of $4.8\,$GHz
to observe each field in the survey.
With $1.5$ hours per field, these observations yielded an RMS flux of
$\sim 25\,\mu$Jy/beam over a $9\pr$ FWHM region with the same pointing center
as a BIMA field.
The positions of all point sources detected with significance $>6\sigma$
within $400\twopr$ of the pointing center have been recorded.
Measured point sources with fluxes corrected for attenuation by the primary beam at $4.8$ GHz
are listed in Tables \ref{tab:vlasrc(1998-2001)} and \ref{tab:vlasrc(2002)}.
 
If the spectra of the point sources are nearly flat or falling,
deep observations with the VLA will identify those that lie near the
noise level in the $28.5$ GHz maps.  However, it is possible that a
radio source with a steeply inverted spectrum may lie below the VLA
detection threshold but would still contribute significantly at $28.5$ GHz.
Advection dominated accretion
flows are thought to be the most common inverted spectrum sources.
They typically have a slowly rising spectrum, with a spectral index of
$0.3$ to $0.4$ (Perna \& DiMatteo, 2000) where the point source flux $S
\propto \nu^{\alpha}$.  Such a shallow spectrum would only provide a
factor of two increase in flux between $4.8$ GHz and $28.5$ GHz;
any source not seen with the VLA would be near the noise level of the 
BIMA observations.

To search for point sources with more steeply inverted spectral
indices, we made VLA observations at $8.0$ GHz of the five BIMA fields 
that most strongly
indicate an excess of anisotropy power. If
radio sources are the dominant contribution to the observed excess
power, it is these fields that are the most likely to be contaminated.
Results of these observations are found in Table
\ref{tab:vlasrc(1998-2001)} and Table \ref{tab:vlasrc(2002)}.  The
$8\,$GHz observations reached a RMS flux density of $20\, \mu$Jy/beam
at the center of the $5^\prime$ FWHM primary beam.  Six of the 13
sources identified in the $4.8$ GHz maps were detected in the $8.0$
GHz maps with Signal-to-Noise ratio (SNR) $> 3$.  The mean spectral index of the detected
sources is found to be $\alpha = -0.4$.
The fluxes of the other seven sources were poorly constrained because those
sources were either too dim or were positioned outside of the $8.0$ GHz primary
beam where the instrument is most sensitive.
The $8\,$GHz images produced no additional point source detections
with SNR$>6$, providing additional evidence against contamination by a
population of radio sources with inverted spectra.

\begin{table*}[!htb]
\caption{\label{tab:vlasrc(1998-2001)}Point Sources and Fluxes Identified from VLA Observations(1998-2001)}
\small
\begin{center}
\begin{tabular}{lccccc}
\hline\hline
\multicolumn{1}{c}{Field} &  $\Delta$ R.A. ($^{\twopr}$) & $\Delta$ DEC ($^{\twopr}$) & 4.8 GHz ($\mu$Jy) &
8 GHz ($\mu$Jy) & 30 GHz ($\mu$Jy)  \\ \hline
BDF4 & $-96.8$ & $\phn 255.7$ & $1230\pm 90.6$ \\
BDF4 & $\phn \phn 72.8$ & $\phn 178.2$ & $\phn 514\pm 54.9$ & &  $468\pm313$ \\
BDF4 & $\phn \phn 99.9$ & $-89.4$ & $\phn 221 \pm 42.3$ & & $90.7\pm225$\\
BDF4 & $-94.9$ & $\phn 268.4$ & $\phn 391 \pm 89.3$ \\
%BDF4 & $-121.0$ & $\phn 356.3$ & $\phn 673 \pm 133$ \\
HDF & $-35.0$ & $-85.0$ & $\phn 832 \pm 56.3$ & & $535\pm185$ \\
HDF & $\phn 255.0$ & $-89.8$ & $1380 \pm 93.0$ \\
HDF & $\phn 178.3$ & $-274.0$ & $1520 \pm 112$ \\
HDF & $\phn 222.5$ & $-86.8$ & $\phn 709 \pm 64.6$ \\
HDF & $\phn \phn 69.1$ & $\phn 334.1$ & $1120 \pm 107$ \\
HDF & $-21.1$ & $\phn \phn 66.1$ & $\phn 190 \pm 36.2$ \\
BDF6 & $-136.5$ & $-283.5$ & $\phn 592 \pm 74.5$  & $-51.3 \pm 226$ \\
BDF7 & $\phn 314.6$ & $\phn \phn 47.4$ & $1554 \pm 80.5$ \\
BDF7 & $\phn 173.8$ & $\phn \phn 97.8$ & $\phn 373 \pm 40.1$ & & $156\pm306$\\
BDF7 & $\phn 253.8$ & $\phn -1.1$ & $\phn 284 \pm 49.5$ \\
BDF8 & $-145.9$ & $-266.1$ & $1381 \pm 94.1$ \\
BDF8 & $\phn \phn 27.6$ & $\phn 280.9$ & $\phn 611 \pm 72.8$ \\
BDF8 & $\phn 302.9$ & $-79.5$ & $\phn 622 \pm 86.1$ \\
BDF9 & $-221.9$ & $-123.7$ & $1500 \pm 78.4$ \\
BDF9 & $-192.7$ & $\phn 215.8$ & $1193 \pm 81.3$ \\
BDF9 & $\phn 245.2$ & $-101.0$ & $1039 \pm 67.8$ \\
BDF10 & $-158.5$ & $-165.6$ & $1670 \pm 63.9$ & & $1610\pm347$\\
BDF10 & $-146.1$ & $-183.9$ & $\phn 320 \pm 44.4$ \\
BDF11 & $\phn \phn 87.7$ & $\phn \phn 77.9$ & $\phn 246 \pm 34.6$ \\
BDF11 & $\phn 342.8$ & $\phn \phn \phn 8.8$ & $\phn 865 \pm 101$ & & $387\pm179$ \\
BDF11 & $\phn \phn 42.5$ & $-11.8$ & $\phn 152 \pm 30.7$ \\
BDF12 & $-241.0$ & $-256.7$ & $1620 \pm 105$ & $-97.4 \pm 368$ \\
BDF12 & $\phn 260.2$ & $\phn 300.5$ & $1191 \pm 129$ & $-1080 \pm 905$ \\
BDF12 & $-137.4$ & $-133.4$ & $\phn 278 \pm 40.5$ & $126 \pm 36.7$ & $165\pm273$ \\
BDF12 & $\phn 170.9$ & $\phn \phn 66.7$ & $\phn 211 \pm 38.7$ & $98.8 \pm 33.9$ & $-151\pm259$ \\
BDF13 & $\phn 181.4$ & $-49.5$ & $\phn 721 \pm 51.2$ & $\phn 531 \pm 50.8$ & $935\pm271$ \\
BDF13 & $-154.0$ & $\phn 299.7$ & $1145 \pm 98.8$ & $126 \pm 427$ \\
BDF13 & $\phn 225.1$ & $-99.9$ & $\phn 317 \pm 56.4$ & $11.4 \pm 106$ \\
\hline
\end{tabular}
\end{center}
\begin{center}
Entries are left blank for
fields not observed at $8\,$GHz.  
For the $30\,$GHz BIMA observations, entries are blank for those point sources which lie outside
the primary beam and are not detected at $>3\sigma$ significance.
\end{center}
\normalsize
\end{table*}
 
\begin{table*}[!htb]
\caption{\label{tab:vlasrc(2002)}Point Sources and Fluxes Identified from VLA Observations(2002)}
\small
\begin{center}
\begin{tabular}{lccccc}
\hline\hline
\multicolumn{1}{c}{Field} &  $\Delta$ R.A. ($^{\twopr}$) & $\Delta$ DEC. ($^{\twopr}$) & 4.8 GHz ($\mu$Jy) &
8 GHz ($\mu$Jy) & 30 GHz ($\mu$Jy) \\ \hline
BDF14 & $\phn 125.1$ & $-166.8$ & $1186 \pm 35.3$ & $1235 \pm 63.1$ \\
BDF14 & $\phn -56.3$ & $\phn \phn 30.3$ & $\phn 226 \pm 24.1$ & $\phn 295 \pm 22.2$ & $-195\pm143$ \\
BDF14 & $\phn \phn 17.6$ & $-183.4$ & $\phn 578 \pm 26.5$ & $\phn 367 \pm 48.9$ & $110\pm244$ \\
BDF15 & $\phn \phn -6.9$ & $-324.4$ & $9390 \pm 78.6$ & & $6390\pm857$\\
BDF15 & $-129.0$ & $\phn 218.7$ & $\phn 637 \pm 42.1$ \\
BDF15 & $\phn \phn 77.4$ & $-282.6$ & $\phn 707 \pm 51.4$ \\
BDF16 & $-157.8$ & $\phn 116.7$ & $1091 \pm 38.3$ & & $212\pm274$\\
BDF16 & $-266.2$ & $\phn 198.5$ & $2553 \pm 74.8$ \\
BDF16 & $-158.4$ & $\phn \phn 33.6$ & $\phn 576 \pm 33.1$ & & $215\pm222$ \\
BDF17 & $\phn \phn 39.9$ & $\phn 169.3$ & $\phn 431 \pm 26.2$ & $\phn 774 \pm 44.4$ & $694\pm228$ \\
BDF18 & $-213.4$ & $\phn 381.6$ & $6902 \pm 145$ \\
BDF18 & $-106.1$ & $\phn -0.7$ & $\phn 465 \pm 25.7$ & & $145\pm168$ \\
BDF18 & $\phn 285.4$ & $\phn \phn \phn 2.8$ & $\phn 988 \pm 49.9$ \\
BDF18 & $\phn 137.7$ & $\phn \phn 93.9$ & $\phn 515 \pm 29.7$ & & $103\pm222$\\
BDF18 & $\phn \phn -9.5$ & $\phn 290.3$ & $1636 \pm 50.5$ \\
BDF18 & $\phn 112.9$ & $\phn 165.9$ & $\phn 266 \pm 32.8$ & & $381\pm276$\\
BDF18 & $\phn -32.2$ & $\phn 105.4$ & $\phn 333 \pm 25.5$ & & $142\pm170$\\
BDF19 & $\phn \phn 31.9$ & $\phn 169.4$ & $\phn 289 \pm 32.8$ & & $-95\pm237$\\
BDF19 & $-169.8$ & $\phn -17.3$ & $\phn 363 \pm 32.6$ & & $-16\pm234$ \\
BDF19 & $-128.7$ & $\phn -76.5$ & $\phn 403 \pm 30.4$ & & $167\pm210$ \\
BDF20 & $\phn 245.8$ & $\phn \phn 15.9$ & $3809 \pm 49.4$ & & $1240\pm400$ \\
BDF20 & $-131.5$ & $\phn 394.5$ & $4957 \pm 136$ \\
BDF20 & $\phn \phn 89.7$ & $\phn 360.3$ & $3272 \pm 96.6$ \\
BDF20 & $-301.9$ & $\phn 149.2$ & $2851 \pm 76.0$ \\
BDF21 & $\phn 277.3$ & $-180.0$ & $2103 \pm 57.5$ \\
BDF21 & $\phn 229.4$ & $\phn \phn 74.1$ & $1187 \pm 34.8$ \\
BDF21 & $-315.9$ & $-224.9$ & $\phn 838 \pm 82.2$ \\
BDF21 & $\phn 115.7$ & $\phn 268.9$ & $\phn 470 \pm 44.2$ \\
\hline
\end{tabular}
\end{center}
\begin{center}
Entries are left blank for
fields not observed at $8\,$GHz.  
For the $30\,$GHz BIMA observations, entries are blank for those point sources which lie outside
the primary beam and are not detected at $>3\sigma$ significance.
\end{center}
\normalsize
\end{table*}

\clearpage

\section{Excess Power Estimate}\label{sec:power}
 
Following the method of D2002, the excess power in the BIMA data is computed assuming that 
the angular power spectrum can be described by either one or two flat band powers.  
Point sources identified with the VLA are tabulated in \S\ref{sec:VLA} and are removed from the data
using the constraint matrix technique described in D2002.
Confidence intervals for the band powers are determined using the integrated likelihood 
also described in D2002.
In Table \ref{tab:powshort},
we show the most likely $\Delta T^2$ and approximate uncertainty computed from visibilities
in the $0.63-1.1\,\kl$ range for each of the 18 fields in the survey.  
Results for the complete data set are included in Table \ref{tab:powall} 
for the cases when the power spectrum is modeled by two bandpowers corresponding to $u$-$v$ 
ranges $0.63-1.1\,\kl$ and $1.1-1.7\,\kl$,
and for a single bin covering the range $0.63-1.7\,\kl$.

In the combined analysis of all 18 fields, we allow for $\Delta T^2 < 0.0$ in determining the most likely
estimate of measured power and confidence intervals.
Excess power corresponding to $\Delta T_1^2 > 0$, was observed with 
$94.7\%$ confidence in the $0.63-1.1\,\kl$ bin.
In the $u$-$v$ range $1.1-1.7\,\kl$, the level of observed power was consistent with zero and 
we found an upper limit $\Delta T_2^2 < 880 \mu{\rm K}^2$ at $95\%$ confidence.
When an analysis is performed combining all data into a single bin, we find an estimate of excess power
excluding zero, $\Delta T^2 > 92.3\%$ confidence.

Window functions for each of these bands are produced from the noise weighted sum of the window
functions for the individual visibilities.
Averaged over all 18 fields, the $0.63-1.1\,\kl$ band has an average value of $\ell_{eff}=5237$ 
with FWHM $\ell=2870$.
The window function for visibilities in the $u$-$v$ range $1.1-1.7\,\kl$ has an average 
value $\ell_{eff}=8748$ with FWHM $\ell=4150$.
For the single band model covering the $u$-$v$ range $0.63-1.7\,\kl$, the window 
function has an average value $\ell_{eff}=6864$ with FWHM $\ell=6800$.

\begin{table}[!htb]
\caption{\label{tab:powshort}Power Estimates for
$0.63-1.1\,\kl$}
\small
\begin{center}
\begin{tabular}{lcc}
\hline\hline
\multicolumn{1}{l}{} & \multicolumn{2}{c}{$\Delta T^2(\mu{\rm K}^2)$} \\
\multicolumn{1}{c}{Field} & Most Likely & $\sigma$  \\\hline
BDF4 & $0.0$ & $600$ \\
HDF &  $0.0$ & $270$ \\
BDF6 & $380$ & $455$  \\
BDF7 & $300$ & $1050$  \\
BDF8 & $0.0$ & $390$ \\
BDF9 & $0.0$ & $590$ \\
BDF10 & $0.0$ & $360$ \\
BDF11 & $50$ & $920$ \\
BDF12 & $1590$ & $1165$  \\
BDF13 & $1480$ & $1260$  \\
BDF14 & $690$ & $920$  \\
BDF15 & $0.0$ & $660$  \\
BDF16 & $300$ & $880$  \\
BDF17 & $1390$ & $1400$  \\
BDF18 & $200$ & $990$ \\
BDF19 & $0.0$ & $900$ \\
BDF20 & $290$ & $910$ \\
BDF21 & $0.0$ & $310$ \\\\ 
\hline
\end{tabular}
\end{center}
\begin{center}
\end{center}
\normalsize
\end{table}
 
The likelihood distribution near the maximum ($\Delta T^2_B={\overline \Delta T^2}_B$) for the data defined by bin $B$ is well
described by a offset log-normal function (Bond, Jaffe, \& Knox, 2000),
\begin{equation}\label{log-normal}
\ln{\mathcal L}({\bf \Delta T^2}) = \ln{\mathcal L}(\overline{{\bf \Delta T^2}})-\frac{1}{2}
\sum\limits_{B}\frac{(Z_B - \overline{Z}_B)^2}{\sigma_B^2}e^{2{\overline Z}_B},
\end{equation}
where the offset log-normal parameters ${\bf Z}$ are defined as
\begin{equation}\label{lognorm}
Z_B = \ln{(\Delta T^2_B + x_B)}.
\end{equation}
The likelihood functions are fit with this model to determine values for the curvature at peak $\sigma_B$,
which represents the uncertainty in the measurement, 
and log-normal offset $x_B$ to the likelihood functions reported in Table \ref{tab:powall}.

\begin{table*}[!htb]
\caption{\label{tab:powall}Power Estimates for Combined Data Sets}
\small
\begin{center}
\begin{tabular}{lcccccc}
\hline\hline
\multicolumn{1}{l}{} & \multicolumn{4}{c}{$\Delta T^2(\mu{\rm K}^2)$}& Likelihood \\
\multicolumn{1}{c}{$u$-$v$ range} &  Most Likely & $68\%$ Confidence & $\sigma_B$ & $x_B$ & $\Delta T^2 > 0$  \\\hline
$0.63-1.1\,\kl$ & $220$ & $100-360$ & $130$ & $625$ & $94.7\%$  \\
$1.1-1.7\,\kl$ & $-40$ & $<420$ & $395$ & $3040$ & $ $  \\
$0.63-1.7\,\kl$ & $170$ & $70-290$ & $110$ & $490$ & $92.3\%$  \\\\
\hline
\end{tabular}
\end{center}
\begin{center}
\end{center}
\normalsize
\end{table*}

\subsection{Point Source Removal}\label{subsec:ptsrc model}
 
As discussed in \S\ref{sec:VLA}, we have adopted a detection threshold of $6\sigma$ for 
identifying point sources in the VLA data.
We measured the effect of eight different point source detection thresholds on the 
measured excess power from the combined analysis of the 18 fields.
We choose point source detection limits in terms of SNR rather than flux to
account for attenuation by the VLA primary beam.
The noise is assumed to have an RMS of $25\,\mu$Jy/beam in all VLA fields.
The results are listed in Table \ref{tab:ptsrc}.
 
\begin{table*}[htb]
\caption{\label{tab:ptsrc}Effect of Point Source Model on $\Delta T_1^2$}
\small
\begin{center}
\begin{tabular}{lccc}
\hline\hline
\multicolumn{1}{l}{} & \multicolumn{1}{c} {} & \multicolumn{2}{c}{$\Delta T_1^2(\mu{\rm K}^2)$} \\
\multicolumn{1}{l}{VLA Detection Limit (SNR)} & Number of Sources & {Most likely} & $\sigma$ \\ \hline
none & $0$ & $430$ & $155$  \\
$>40$ & $7$ & $350$ & $150$  \\
$>20$ & $27$ & $290$ & $140$ \\
$>12$ & $45$ & $260$ & $130$  \\
$>8$ & $58$ & $240$ & $135$ \\
$>6$ & $62$ & $220$ & $130$ \\
$>5$ & $98$ & $210$ & $135$ \\
$>4$ & $168$ & $250$ & $150$ \\
\hline
\end{tabular}
\end{center}
\normalsize
\end{table*}

In the case for which no point sources are removed, the most likely value of $\Delta T_1^2 = 430 \, \mu{\rm K}^2$
appears significantly elevated due to contamination by radio point sources. 
The measured excess power drops from $430 \, \mu{\rm K}^2$ in the case of no point 
source constraints to a broad minimum of $\sim 200 - 250 \, \mu{\rm K}^2$ for removal of 
sources detected with less than $8 \sigma$ significance.
As the detection threshold decreases, the removal of point sources will begin to remove a significant 
number of degrees of freedom from the analysis, resulting in increased uncertainty in the 
measurement of power.  
For example, a detection threshold for point sources of $4\sigma$
removes three times as many sources from the data as a detection threshold of $8-12 \sigma$.
The uncertainty in the measurement begins to significantly increase with a point source 
detection threshold less than $6 \sigma$ while the subtraction of the additional sources have
no effect on the observed excess power. 
Therefore, we adopt $6 \sigma$ as the point source detection threshold. 

While the contribution to anisotropy from point sources is expected to scale as
$\ell^2$, the estimate of excess power is
consistent with zero on finer angular scales as
shown in Table \ref{tab:powall}.
It should also be noted that no additional point sources were found
in the 8 GHz VLA observations.
These observations reinforce the conclusion that 
the contribution of point sources to the observed excess power is well constrained in the 
power spectrum analysis.

\subsection{Systematic Tests}\label{subsec:jacknife}
\begin{figure*}[!htb]
\centerline{
  \includegraphics[scale=0.45]{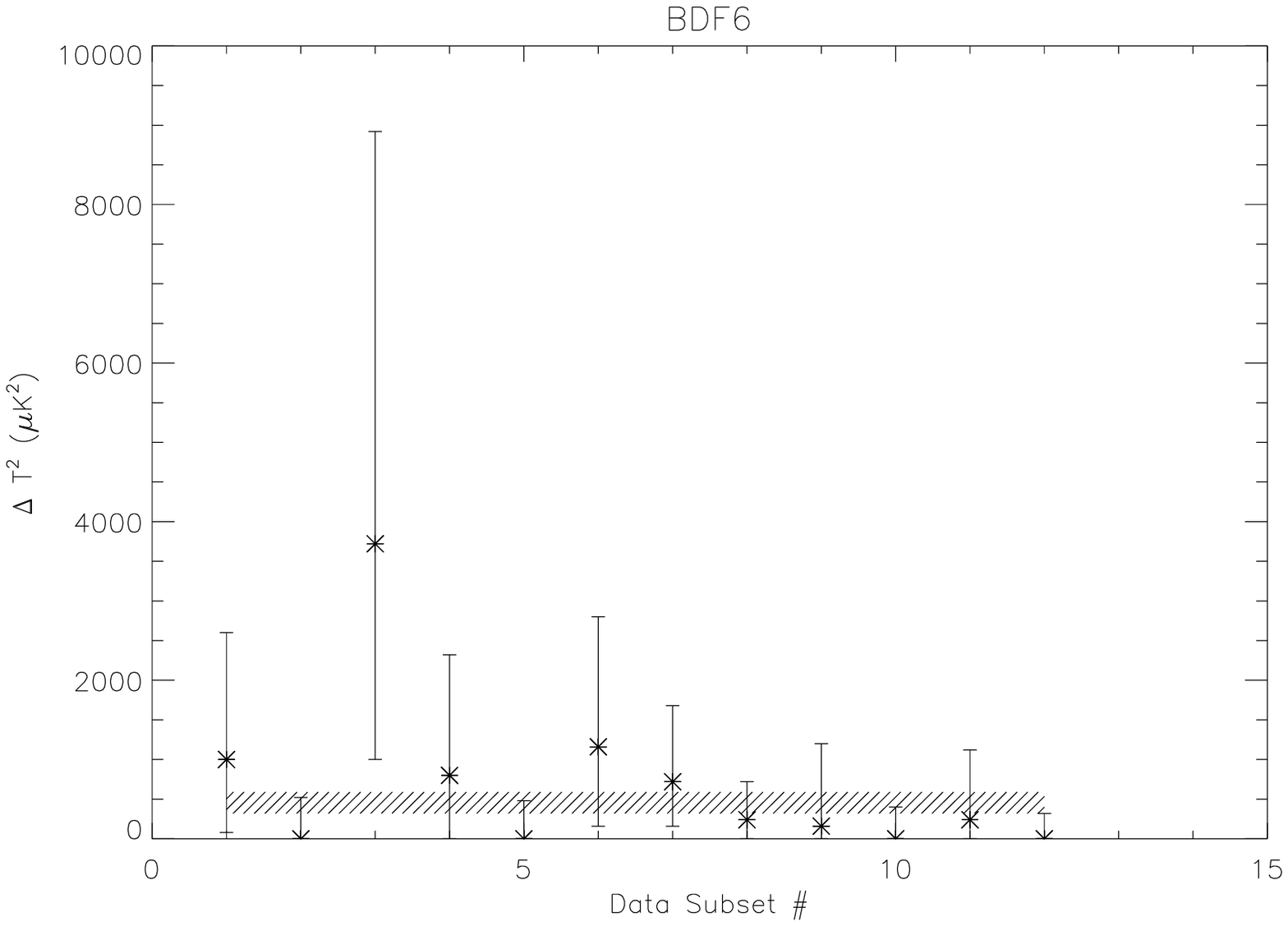}
  \includegraphics[scale=0.45]{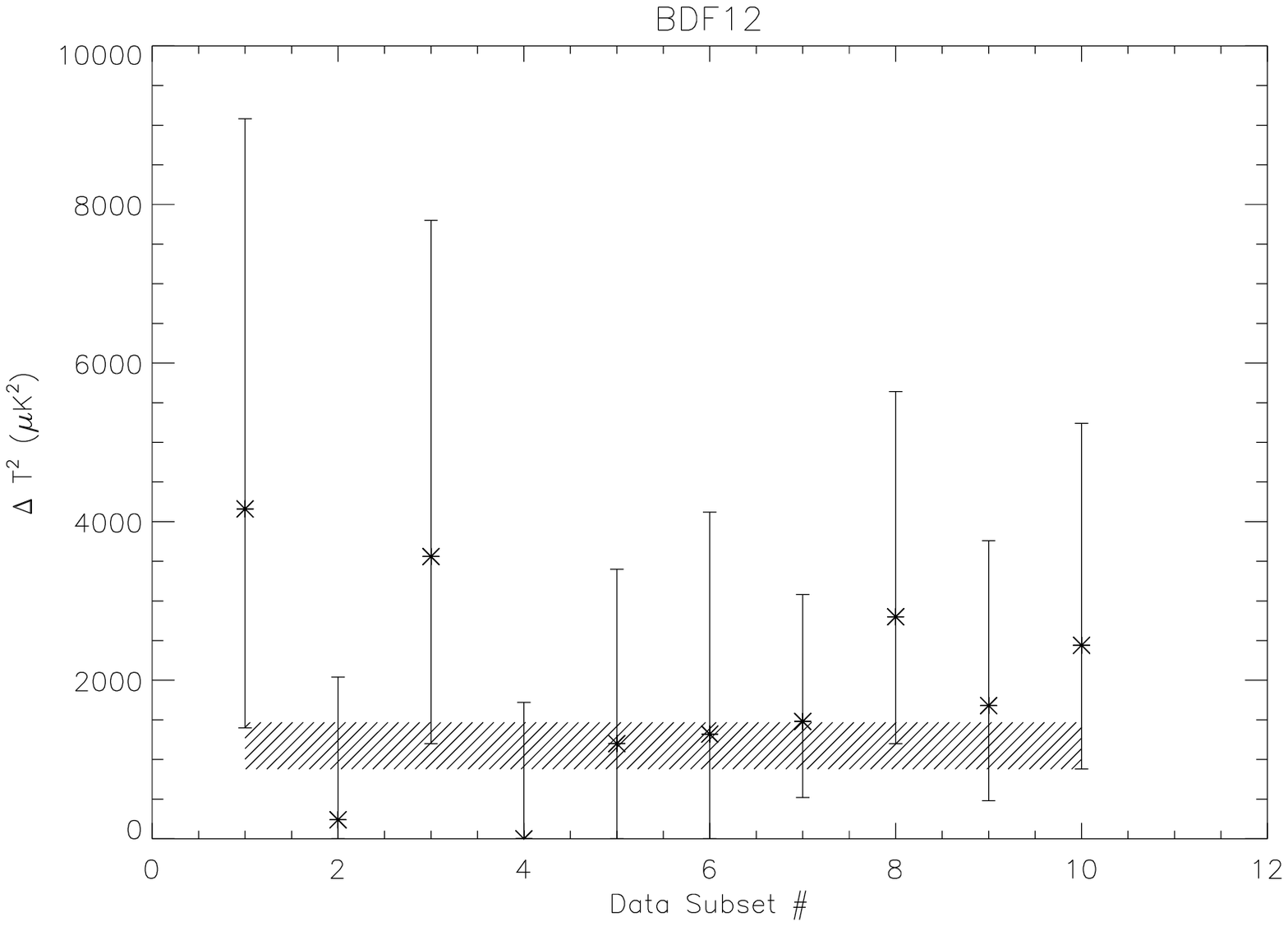}
}
\centerline{
  \includegraphics[scale=0.45]{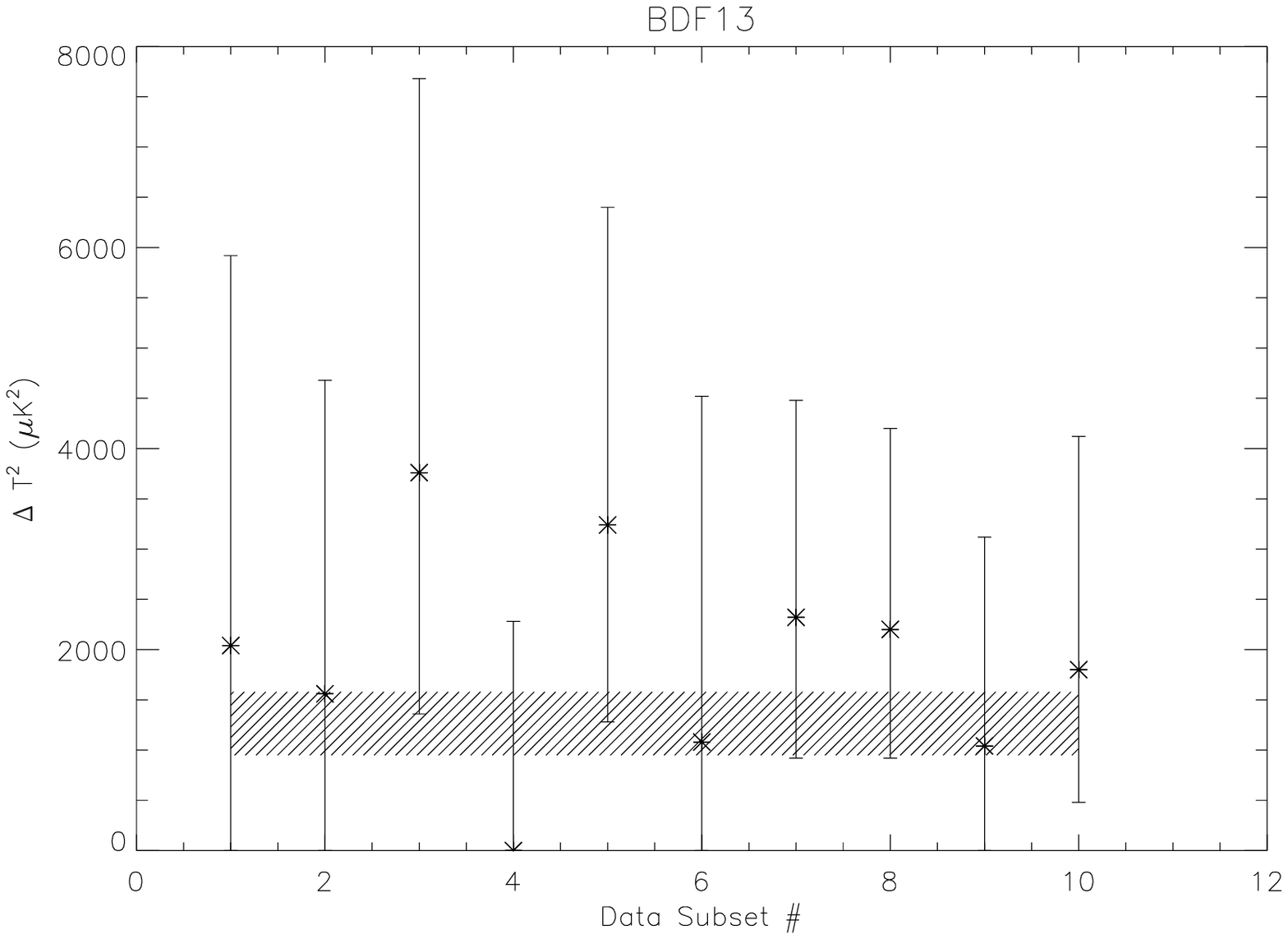}
  \includegraphics[scale=0.45]{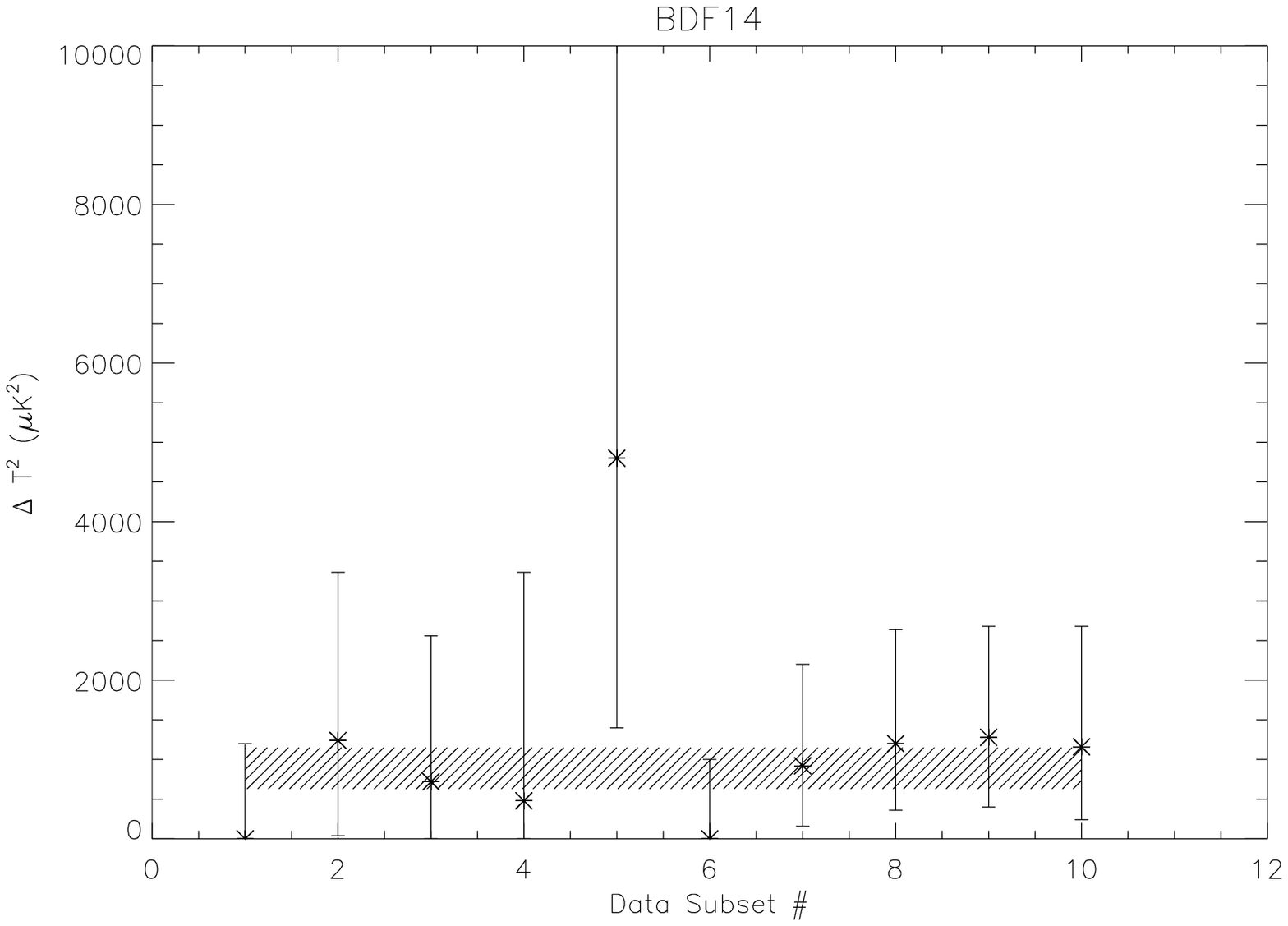}
}
\centerline{
  \includegraphics[scale=0.45]{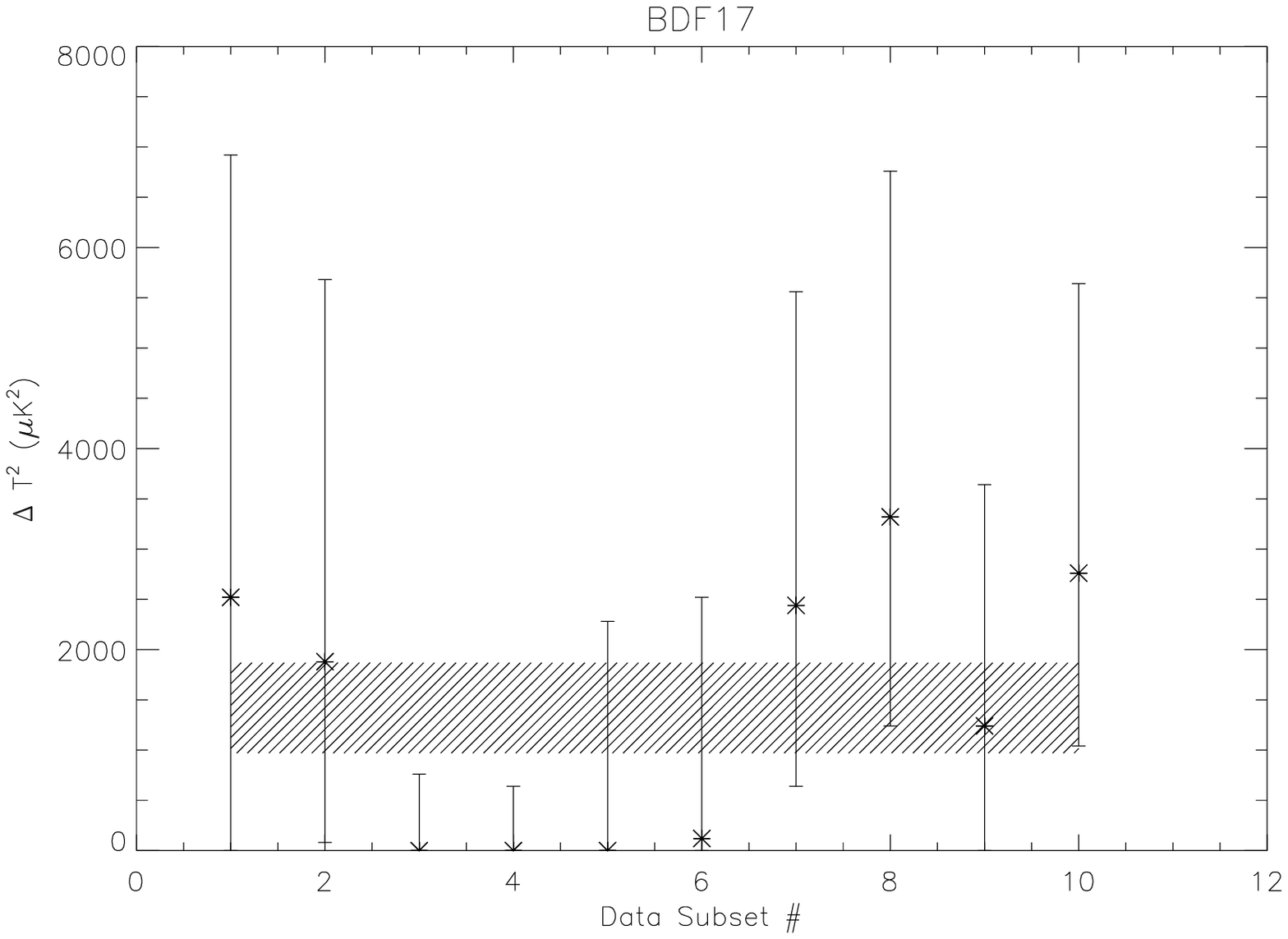}
}
\caption[Results of Systematic Tests]
{Band power estimates and confidence intervals for each systematic test
in those fields found to have significant levels of excess power.
The shaded region represents the $68\%$ confidence intervals
for the full data set of each field.
}
\label{fig:systest}
\end{figure*}

We performed tests for systematic errors in all fields identified as
having significant excess power.
This analysis was limited to visibility data in the $u$-$v$ range
$0.63-1.1\,\kl$ described by $\Delta T_1^2$ where the most significant
detection of excess power occurs.  The modeled power in the second bin is
fixed at $\Delta T_2^2=0$ for all tests described in this section.  We
searched for systematic errors by repeating the power spectrum analysis
after splitting the data into a number of subsets.  The first set of
tests was designed to search for systematic contamination that changes
with time.  Relative to a source on the celestial sphere, terrestrial
sources will appear to move rapidly over the course of a single
observation, while the sun or moon will vary in position by many
degrees over the course of a typical month long observation.  We
searched for such signals by analyzing the data after dividing it into
subsets corresponding to the first, second, and third sections of
both observation tracks and the period of observation.  For the field BDF6,
the only field observed in multiple years that was found to have
a significant level of excess power, we also compared the results of observations
taken in 1998 and 2000.  Instrumental effects that manifest
themselves as spurious signals on a given telescope or baseline are
also a potential source of systematic error.  We searched for such
effects by breaking the data into subsets of four and five telescopes
and looking for antenna based systematic errors.  For a test of
baseline based systematic errors, we created east-west and north-south
baseline subsets.  The different data and instrument subsets used in 
the search for systematic errors are listed in Table \ref{tab:jack-knife}.  
The results for the application of these tests applied to fields BDF6, 
BDF12, BDF13, BDF14, and BDF17 are shown in Figure \ref{fig:systest}.
 
\begin{table}[!htb]
\caption{\label{tab:jack-knife}Cuts Used in Systematic Tests}
\small
\begin{center}
\begin{tabular}{lc}
\hline\hline
\multicolumn{1}{l}{Test Number} & {Data Subset} \\ \hline
1 & First Four Hours UT \\
2 & Middle Three Hours UT \\
3 & Final Four Hours UT \\
4 & First Third of Observation Dates \\
5 & Middle Third of Observation Dates  \\
6 & Final Third of Observation Dates  \\
7 & Baselines from Subarray of four Telescopes \\
8 & Baselines from Subarray of five Telescopes \\
9 & East-West Baselines \\
10 &  North-South Baselines \\
11 &  First Year of Observation (BDF6 only) \\
12 &  Second Year of Observation (BDF6 only) \\
\hline
\end{tabular}
\end{center}
\normalsize
\end{table}
 
\clearpage

If a systematic error was associated with one of these subsets,
then the level of excess power in that subset should increase.
However, as can be seen in Figure~\ref{fig:systest}, we found
approximately the same level of excess power in each subset.  Of the
fifty-two subsets, only the final four hours of UT in BDF6 (third data
point in Figure~\ref{fig:systest}) and the middle third of observation
dates for BDF14 (fifth data point in Figure~\ref{fig:systest}) are
found to have estimates of excess power and $68\%$ confidence
intervals that lie above the $68\%$ confidence limits of the full
analysis.
Therefore, we consider these two subsets to be the most likely to be systematically biased.
To test the possible contribution from systematic errors in these subsets to the measured 
excess power, we repeated the analysis on the 18 combined fields after removing these two subsets.
We found an estimate of $\Delta T^2 = 180^{+140}_{-120}\,\mu$K$^2$, not significantly different
than the reported value of $\Delta T^2 = 220^{+140}_{-120}\,\mu$K$^2$ for the entire survey.
 
\subsection{Correlated Signal Between Independent Fields}\label{subsec:correlation}
 
It is possible that a hardware malfunction could create a
systematic false correlation that is constant or changes slowly with
sky position.  To test for this, we combined the raw
visibilities from observations of all fields taken in a
single summer, as if all the data came from a single pointing.
If the fields contain random noise or independent sky
signal, the power in the combined fields should decrease significantly when the
visibilities are averaged.
A false detection caused by correlations introduced in the hardware or other local effects
might, depending on its stability, enhance the excess power in the anisotropy measurement 
when the independent observations are combined.
The results of the analysis of the combined data sets,
listed in Table \ref{tab:indep}, are  consistent with instrumental
noise at $68\%$ confidence, as expected for non-correlated independent observations.

To test for false correlations that vary slowly across the sky, we observed
a set of fields offset by $2^{\prime}$ in RA from fields that had been previously
observed.
The fields BDF6, BDF12, BDF13, BDF14, and BDF17 have the most significant levels of excess power
of the individual fields in the survey.
We have neighboring fields for all of these except for BDF13.
The raw visibilities of the neighboring fields were combined and then analyzed for excess power;
the results of this analysis are shown in Table \ref{tab:indep}.
In all cases, the observed anisotropy power decreases as we would expect for
the combination of visibilities from independent patches of sky.
Therefore, we conclude  that there does not appear to be any false correlation
that is constant or slowly varying with sky position.
In addition, there does not appear to be any correlation between excess power and proximity to the sun, none of
the fields analyzed for systematic error were near the sun's location of $11-12$ hr R.A. during the summer months.
Overall, we find no evidence that our results are biased by systematic effects
or astrophysical contamination.

\begin{table}[!htb]
\caption{\label{tab:indep}Results of Combining Independent Observations}
\small
\begin{center}
\begin{tabular}{lcc}
\hline\hline
\multicolumn{1}{l}{} & \multicolumn{2}{c}{$\Delta T_1^2(\mu{\rm K}^2)$} \\
\multicolumn{1}{l}{Dataset} & {Most likely} & $\sigma$ \\ \hline
1998 data  & $60$ & $220$  \\
2000 data & $0$ & $220$  \\
2001 data & $40$ & $180$ \\
2002 data & $0$ & $40$  \\
BDF12 & $1590$ & $1165$ \\
BDF19 & $0.0$ & $900$ \\
BDF12/2+BDF19/2 & $560$ & $740$ \\
BDF6 & $380$ & $1050$ \\
BDF17 & $1390$ & $1400$ \\
BDF6/2+BDF17/2 & $640$ & $580$ \\
BDF4 & $0$ & $600$ \\
BDF14 & $690$ & $920$ \\
BDF4/2+BDF14/2 & $120$ & $220$ \\
\hline
\end{tabular}
\end{center}
\begin{center}
\end{center}
\normalsize
\end{table}

\subsection{Constraints on $\sigma_8$} \label{subsec:sig8}

As was described in the introduction, the most likely astrophysical source of the excess power 
observed in the survey is expected to be CMB anisotropy arising from the SZ effect in clusters 
of galaxies.
Assuming that the observed excess power is entirely due to the SZ effect, we use
a publicly available archive of N-body simulations \footnote{Data available at
http://pac1.berkeley.edu/tSZ/}
(Schulz \& White, 2003) to add the expected signal from the SZ
effect to Monte Carlo noise realizations.  These simulations
predict the evolution of mass from $z=60$ to present in a cosmological
model described by $\Omega_M=0.3$, $\Omega_{\Lambda}=0.7$,
$\Omega_bh^2=0.02$, $h=0.7$, $n=1$ and $\sigma_8=1.0$.
Only dark matter is included in these simulations, 
and the baryons contributing to the SZ signal are added assuming 
that the gas closely traces the dark matter.  
Independent regions of the
N-body simulations are multiplied by the $6.6\pr$ FWHM primary beam and
transformed into the $u$-$v$ plane with the same $u$-$v$ sampling as the real BIMA data.  
Noise is added to each $u$-$v$ point with a variance determined from the observed visibilities.
Therefore, each simulated observation has $u$-$v$ coverage
and noise characteristics identical to the real BIMA observation of
each field.
The analysis of one hundred realizations of the BIMA survey with unique instrumental noise 
and simulated SZ sky
resulted in an excess power of $\Delta T^2 = 216 \pm 190\,\mu$K$^2$ at $68\%$ confidence; 
this is remarkably close to the level of excess power found in the BIMA survey.

\begin{figure}[htb]{}
\plotone{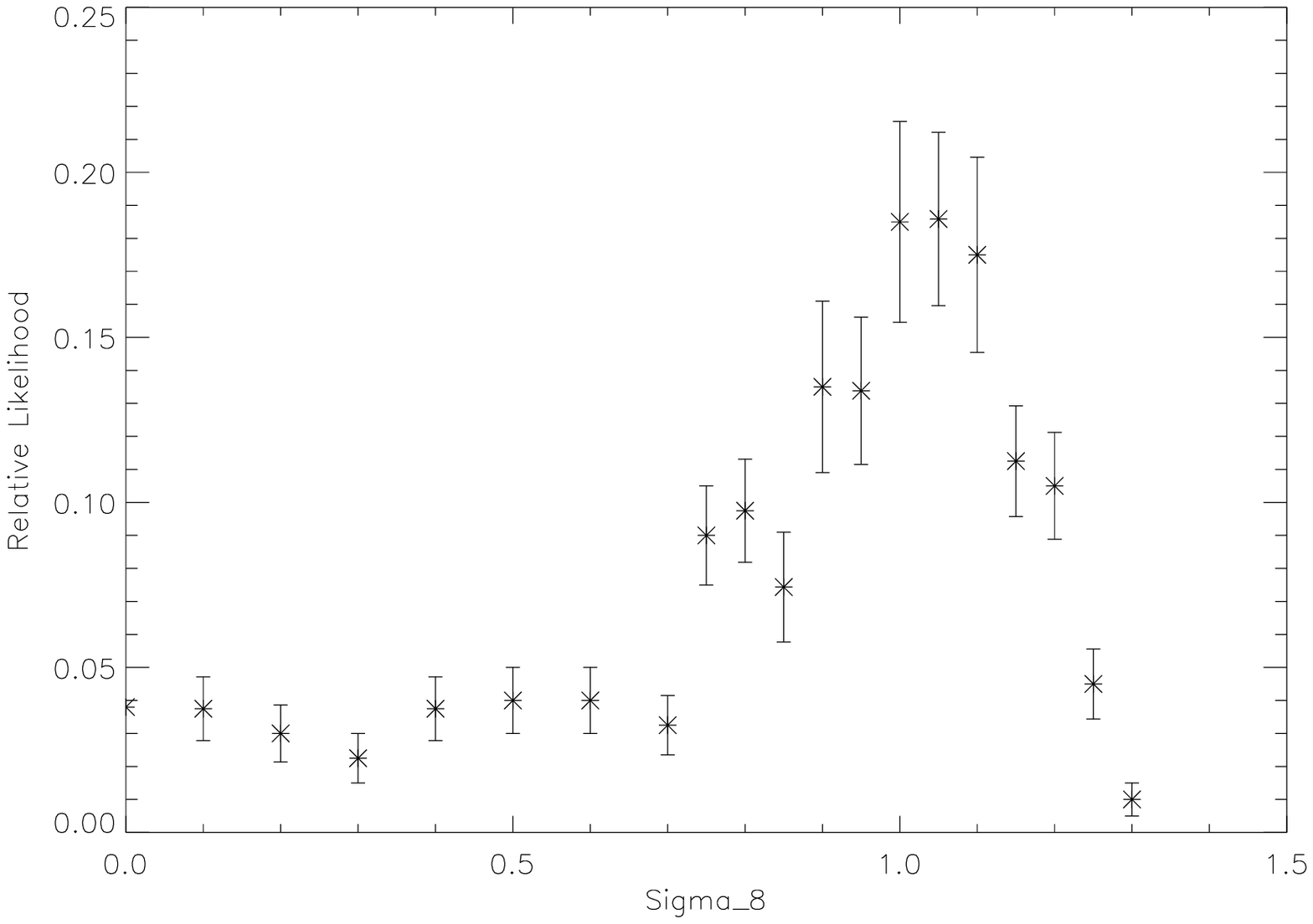}
\caption[Relative likelihood for $\sigma_8$]{
Relative likelihood of measuring the observed excess power as a function of $\sigma_8$.
The uncertainty at each point is determined by the Poisson statistics for the number of 
events in that bin.
}
\label{fig:sig8_like}
\end{figure}

Komatsu and Seljak (2002) demonstrate that the amplitude of the SZ power spectrum has a strong 
dependence on $\sigma_8$, with $\Delta T^2 \propto \sigma_8^7$ for $\sigma_8$ near unity.
We scale the simulated SZ images by $\sigma_8^{7/2}$ to produce skies corresponding to different values of
$\sigma_8$. 
These images are transformed to the $u-v$ plane, combined with the simulated instrumental noise, and used to compute the likelihood
of each value of $\sigma_8$ resulting in the observed excess power, 
which we approximate as $\Delta T^2_1= 180-260\mu$K$^2$.
The relative likelihood that simulations with a given value of $\sigma_8$ will reproduce the 
observed excess power is determined from 10,000 realizations of the BIMA survey with $\sigma_8$ 
ranging from $0.0$ to $1.5$
The resulting relative likelihood shown in Figure~\ref{fig:sig8_like} includes
contributions to the uncertainty from both noise in the measurement and 
sample variance due the non Gaussian nature of the SZ signal and the small patch of sky surveyed.  
Assuming an additional $10\%$ uncertainty in the simulations 
(Komatsu \& Seljak, 2002, Goldstein \ea, 2003),
we find $\sigma_8 = 1.03^{+0.20}_{-0.29}$ at $68\%$ confidence
and $\sigma_8 = 1.03^{+0.30}_{-0.96}$ at $95\%$ confidence.

\section{Image Analysis}\label{sec:images}

Analysis of the complete BIMA survey results in a detection of excess power 
with nearly $95\%$ confidence.    
However, with the power spectrum analysis, we are unable to determine if the
excess power is caused by sources with negative flux as would be expected from the SZ
effect in galaxy clusters at 30 GHz.
The statistics of the pixel flux values in the images contain additional information that 
can in principle constrain possible sources of the excess power.

\begin{figure}[!htb]
\centerline{
  \includegraphics[scale=0.3]{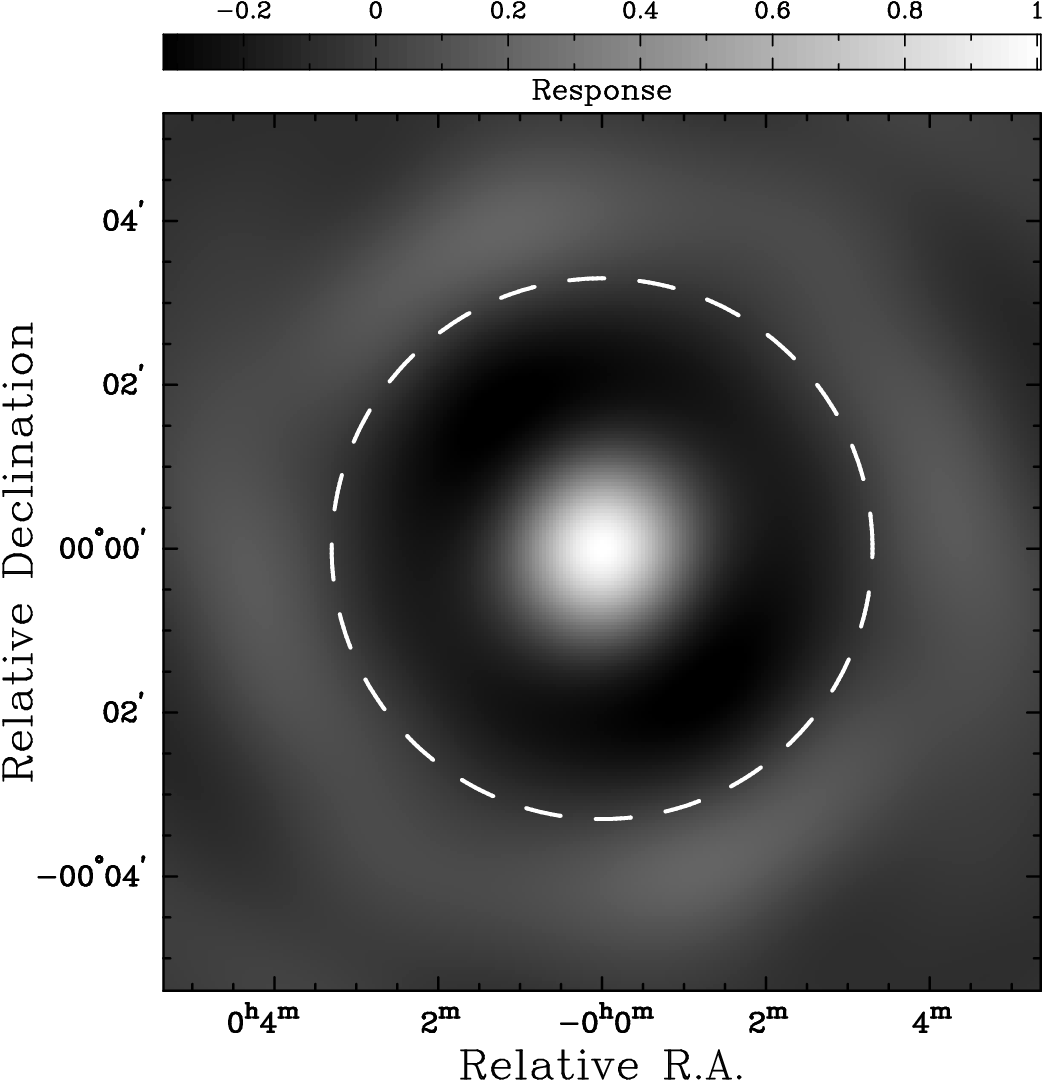}
}
\caption[Typical beam]{
\\
Typical synthesized beam created from the $u$-$v$ coverage used in the analysis of image statistics.
The dotted line shows the half-power radius of the primary beam. 
}
\label{fig:beam}
\end{figure}

Images are reconstructed directly from the visibility data for each
field in the BIMA survey using the DIFMAP software package (Pearson et
al.\ 1994).  Positions determined from the VLA data are used to model
and remove point sources  ($>6\sigma$) in the $u$-$v$ plane using baselines
with $u$-$v$ radius $> 1.5 \, \kl$ ($\sim 50\%$ of the visibilities).
BIMA fields are imaged at a resolution of $7.5\twopr$ per pixel after applying a Gaussian taper with a
half-power radius of $1.0 \,\kl$ to the visibility data to
maximize brightness sensitivity.
These mapping parameters are used
when presenting images of bright SZ clusters observed with the BIMA array (i.e. Laroque et al. 2003)
and produce a synthesized beam which is well matched to cluster scales.
A map of a typical synthesized beam can be found in Figure \ref{fig:beam}.
The response of the first negative sidelobe
is a factor of four lower than response at the center of the beam. 

The typical RMS noise is $110\,\mu$Jy/beam for an image with $90\twopr$ FWHM synthesized beam.
The RMS for each image listed in Table~\ref{tab:properties} are computed
directly from the noise properties of the images.
The RMS temperature estimates correspond to the synthesized beamsize computed for the
$u$-$v$ coverage of each experiment.

\begin{table}[!htb]
\caption{\label{tab:properties}Properties of Each Image Used in Analysis of Image Statistics}
\small
\begin{center}
\begin{tabular}{lccc}
\hline\hline
\multicolumn{1}{l}{} & Synthesized & RMS & RMS  \\
\multicolumn{1}{c}{Field} &  Beamsize($^{\twopr}$) & {($\mu$Jy$\,${\rm beam}$^{-1}$)} & ($\mu$K) \\\hline
BDF4 &  $90.5 \times 105.1$ & $107.1$ & $16.9$  \\
HDF &  $91.4 \times 95.9$ & $113.7$  & $19.5$  \\
BDF6 &  $90.7 \times 97.5$ & $92.4$ & $15.7$  \\
BDF7 &  $91.2 \times 98.2$ & $105.4$ & $17.6$  \\
BDF8 &  $87.9 \times 90.4$ & $109.9$ & $20.7$  \\
BDF9 &  $88.0 \times 91.7$ & $111.4$ & $20.7$  \\
BDF10 &  $88.0 \times 90.3$ & $109.4$ & $20.7$ \\
BDF11 &  $89.0 \times 90.4$ & $109.8$ & $20.5$ \\
BDF12 &  $88.7 \times 92.0$ & $112.3$ & $20.6$ \\
BDF13 &  $89.6 \times 91.3$ & $113.1$ & $20.7$ \\
BDF14 &  $86.6 \times 89.9$ & $106.9$ & $20.6$ \\
BDF15 &  $87.2 \times 90.5$ & $109.1$ & $20.7$ \\
BDF16 &  $87.4 \times 91.5$ & $108.7$ & $20.4$ \\
BDF17 &  $86.0 \times 90.3$ & $105.2$ & $20.3$ \\
BDF18 &  $86.3 \times 91.7$ & $107.7$ & $20.4$ \\
BDF19 &  $88.0 \times 90.7$ & $109.5$ & $20.6$ \\
BDF20 &  $86.8 \times 91.6$ & $108.4$ & $20.4$ \\
BDF21 &  $86.8 \times 91.8$ & $106.7$ & $20.1$ \\\\
\hline
\end{tabular}
\end{center}
\begin{center}
\end{center}
\normalsize
\end{table}

To provide a visual representation
of the data used in the analysis, images of the BIMA fields
are reproduced in Figure \ref{fig:BIMA Image
statistics(1)}.  
The dashed circle in each image represents the radius at which the primary beam
attenuates the sky signal by a factor of two relative to the pointing center.
Regions lying far outside the primary beam
can be used to compute the level of instrumental noise in the map.
 
\begin{figure*}[!h]
\centerline{
  \includegraphics[scale=0.3]{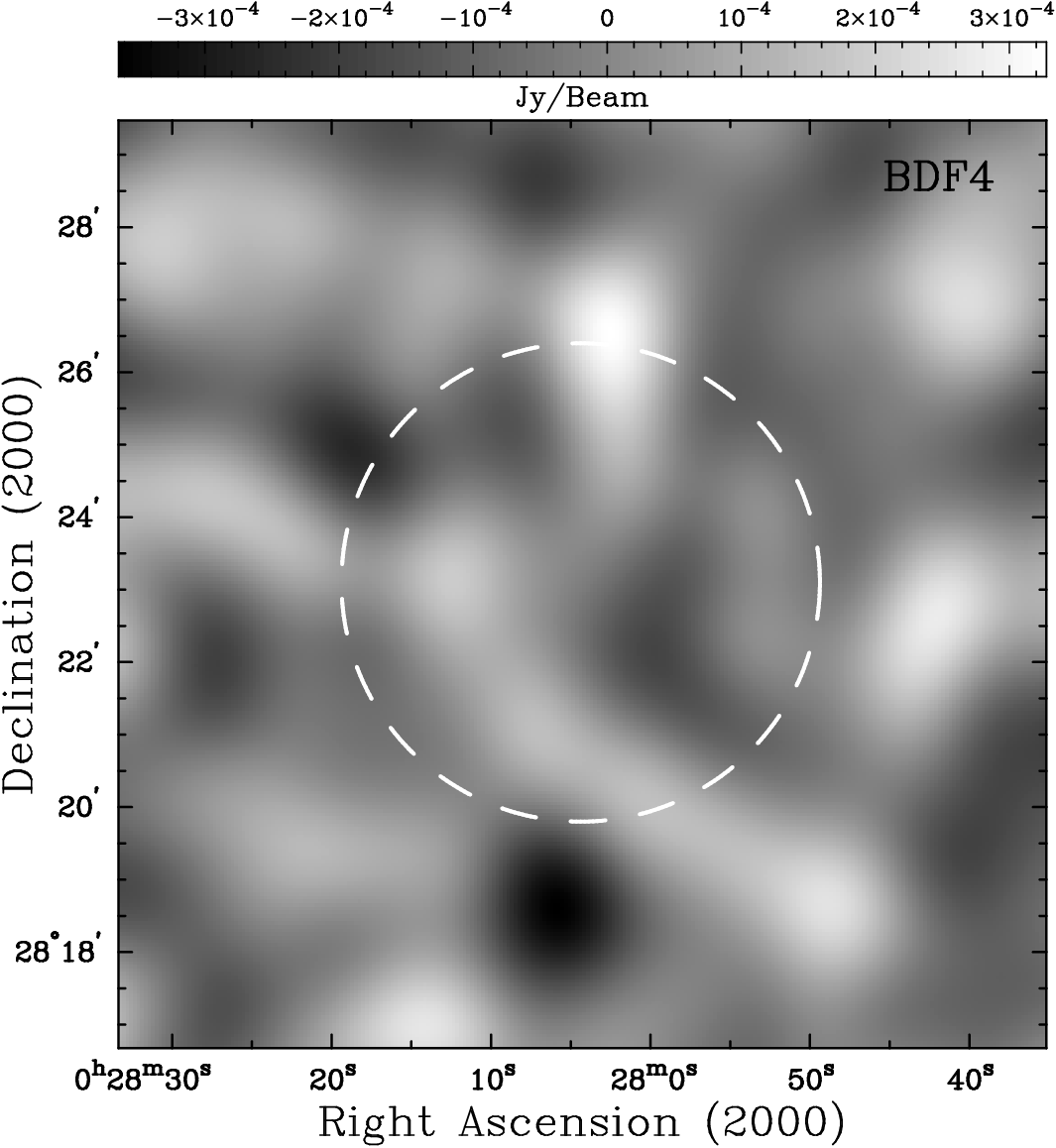}
  \includegraphics[scale=0.3]{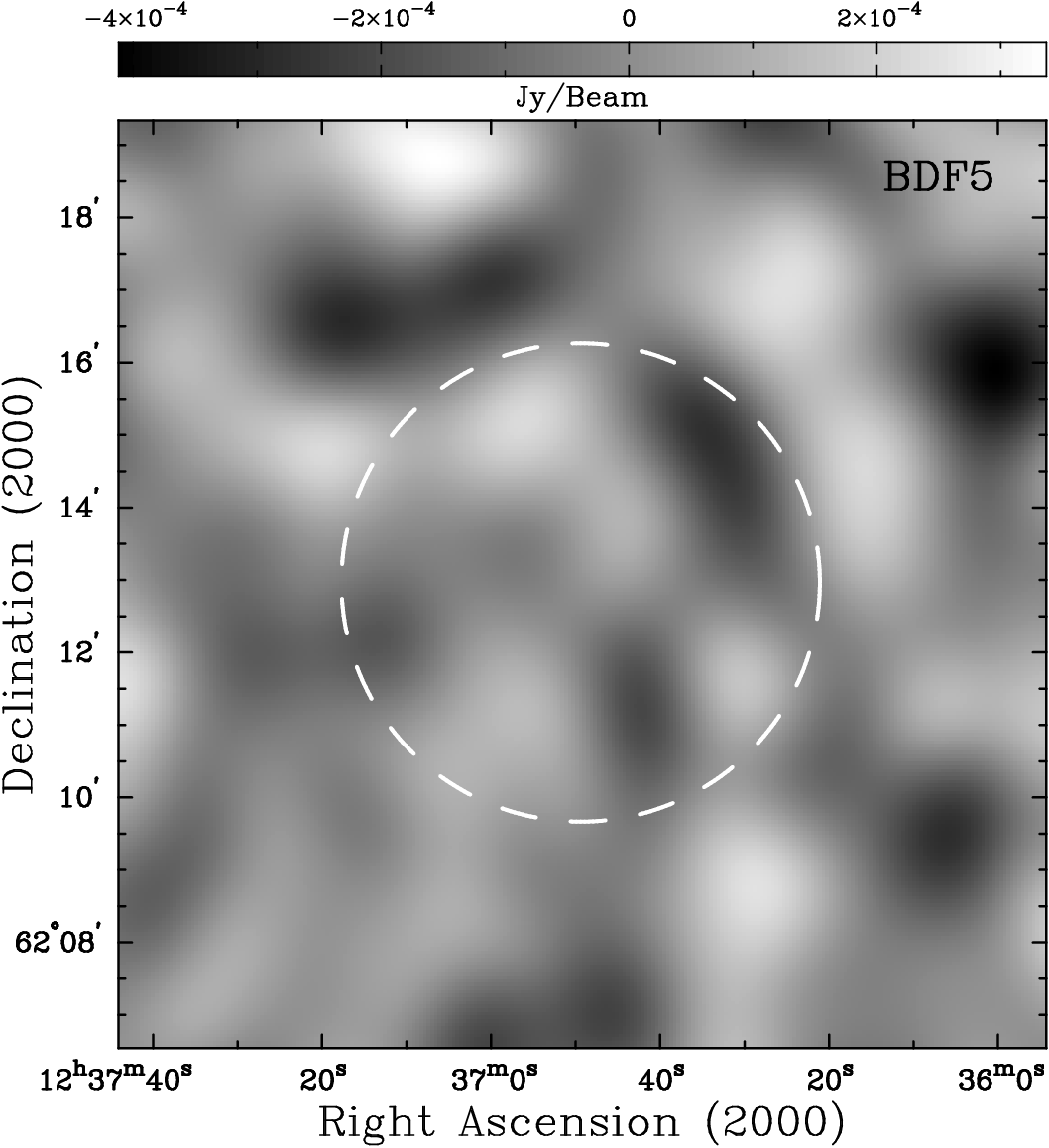}
  \includegraphics[scale=0.3]{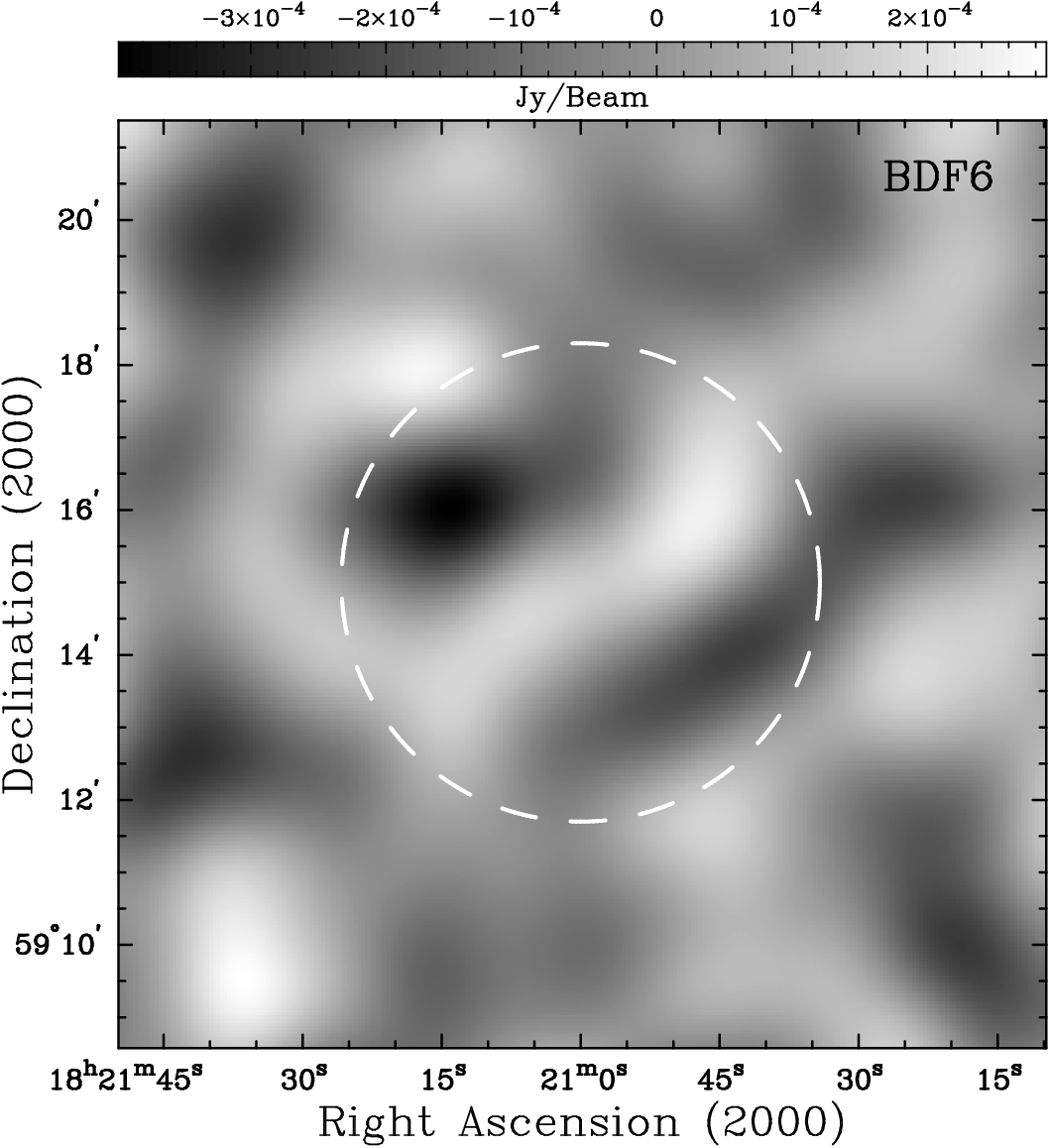}
}
\centerline{
  \includegraphics[scale=0.3]{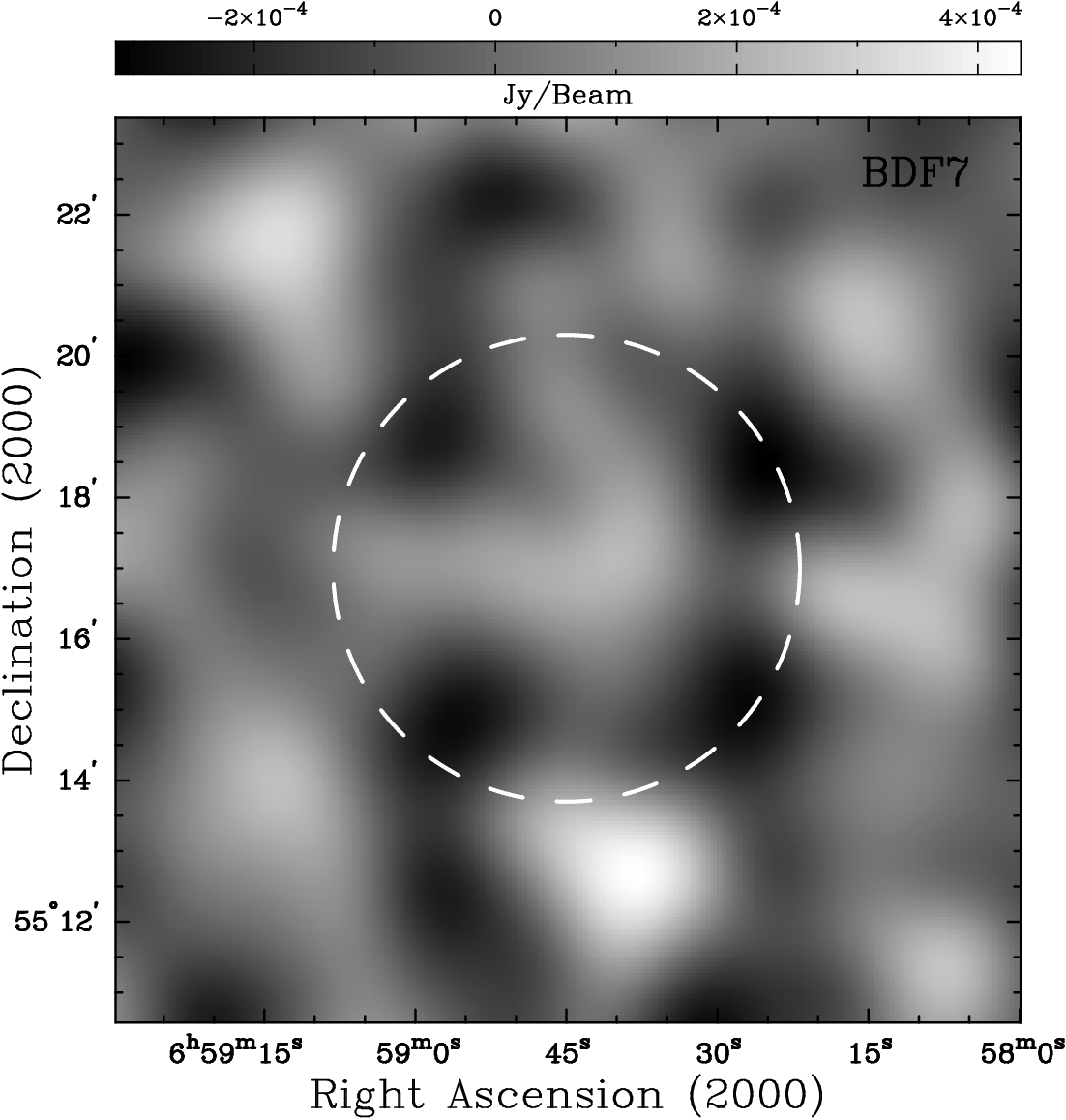}
  \includegraphics[scale=0.3]{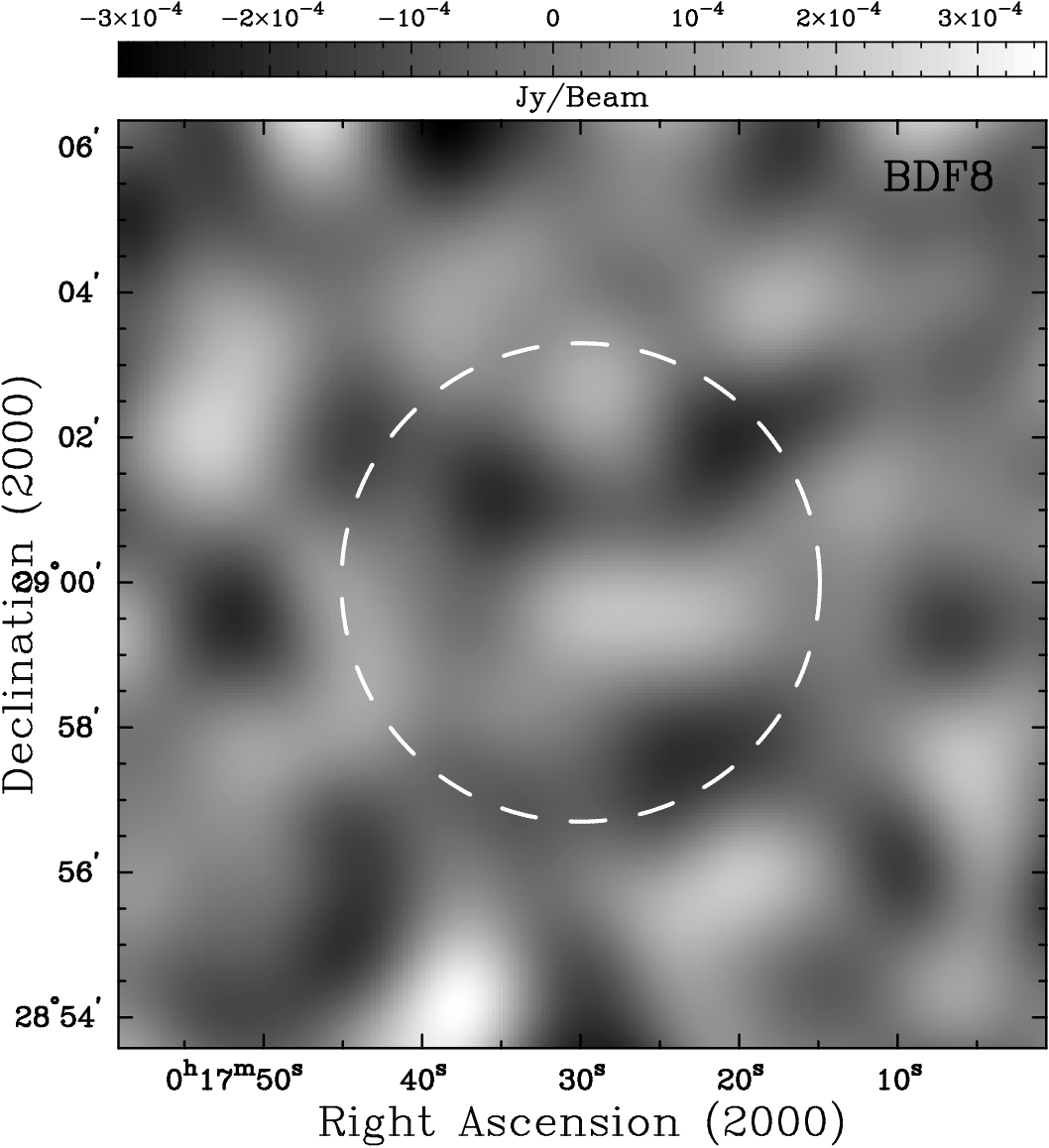}
  \includegraphics[scale=0.3]{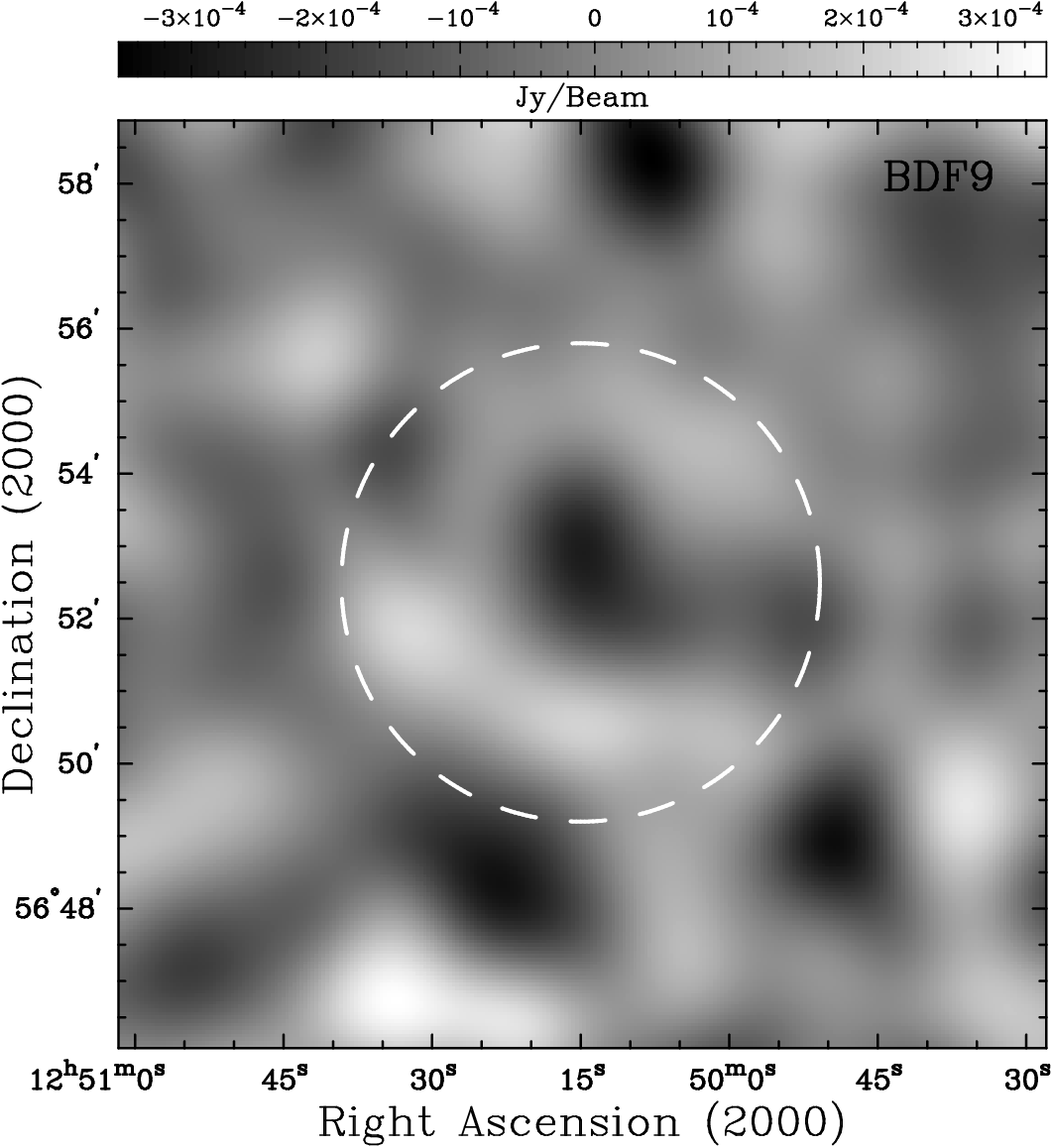}
}
\centerline{
  \includegraphics[scale=0.3]{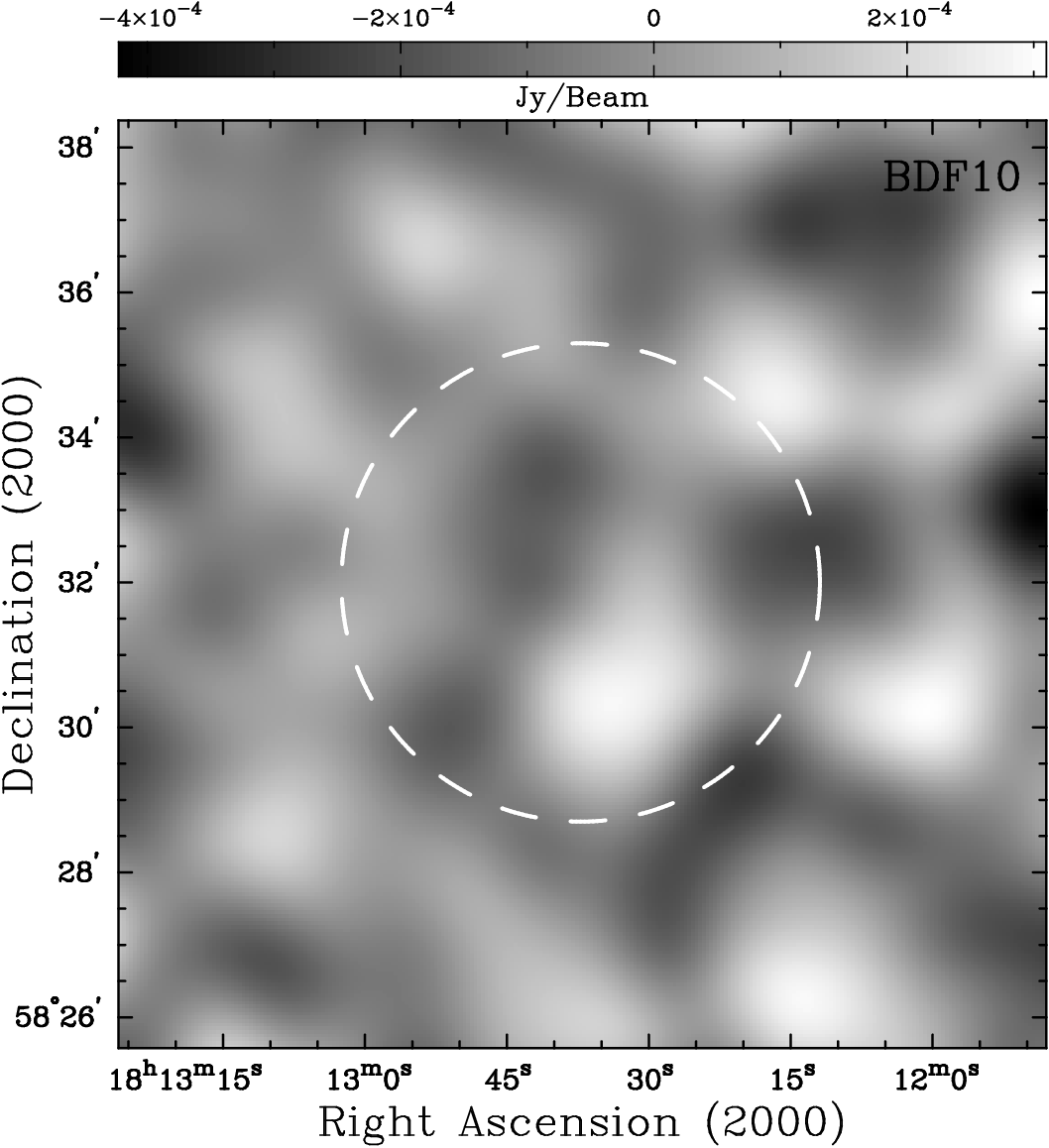}
  \includegraphics[scale=0.3]{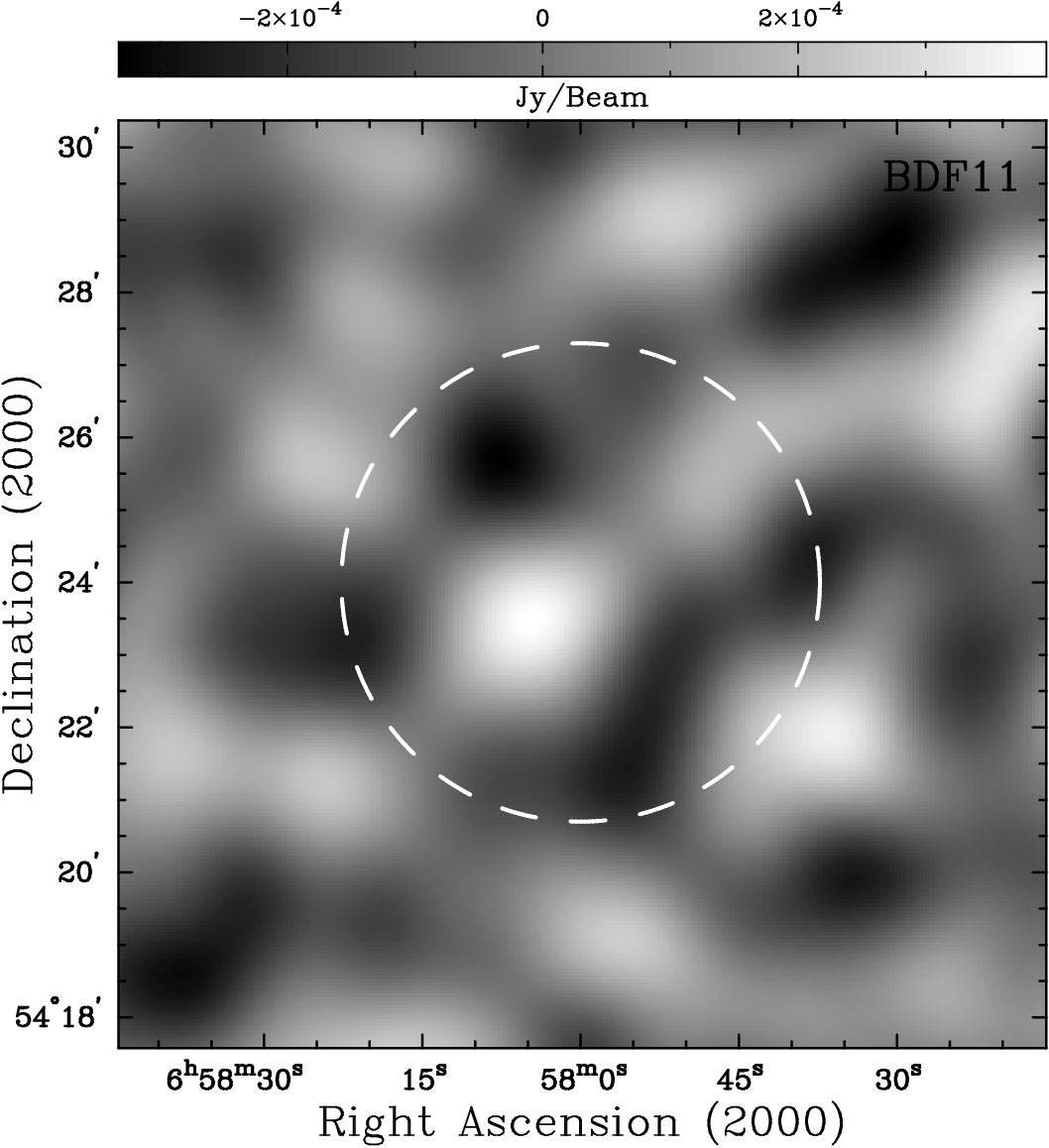}
  \includegraphics[scale=0.3]{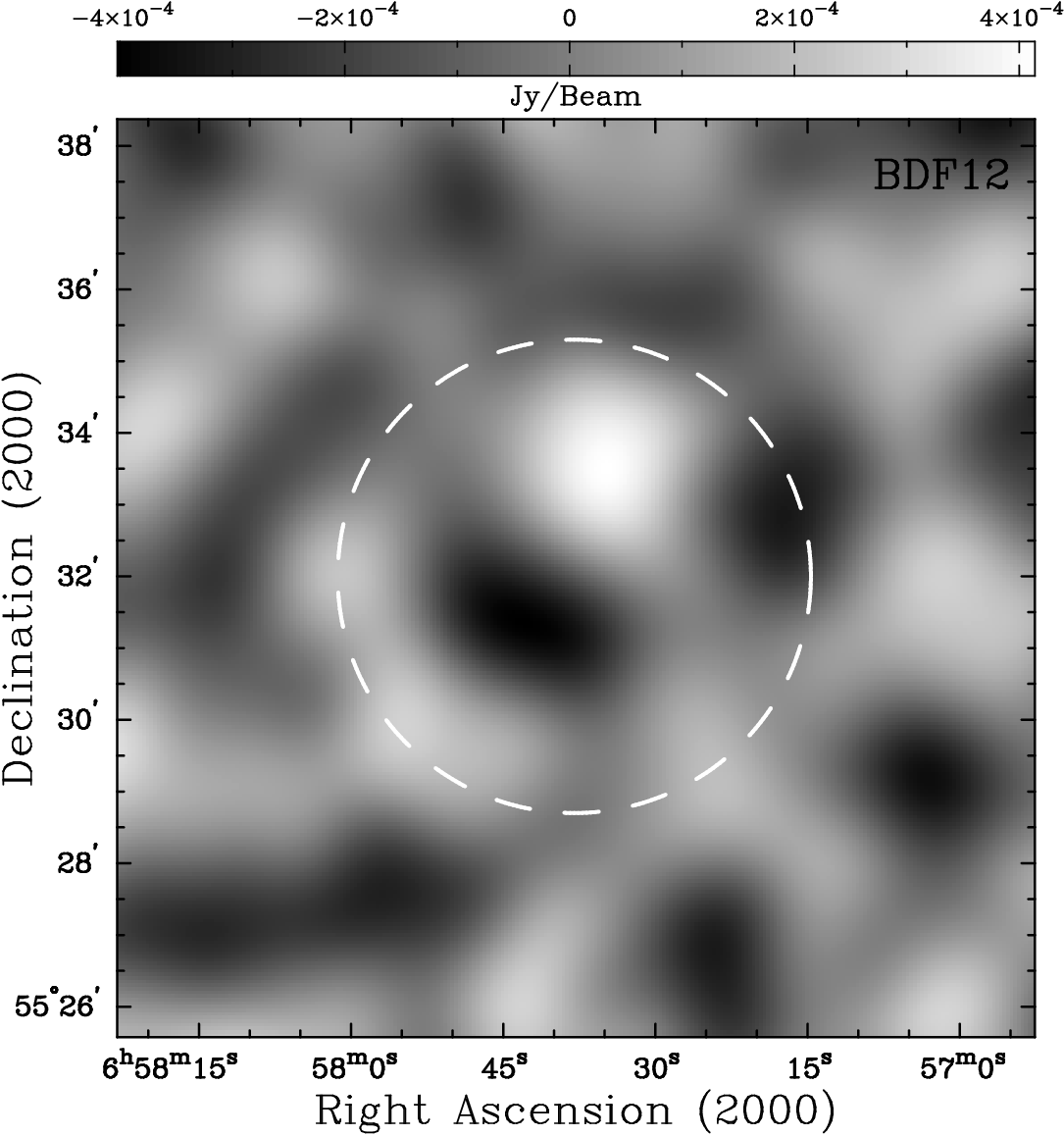}
}
\caption[BIMA Images (BDF4-BDF12)]{BIMA Images after applying
a Gaussian taper with a half-power radius of $1.0 \kl$ to the
visibility data.  Point sources identified in the VLA images have been
modeled from the long baselines ($u$-$v > 1.5 \kl$) and removed
directly from the visibility data.  Dashed lines correspond to the
half power radius of the primary beam.
}
\label{fig:BIMA Image statistics(1)}
\end{figure*}
 
\begin{figure*}[!h]
\centerline{
  \includegraphics[scale=0.3]{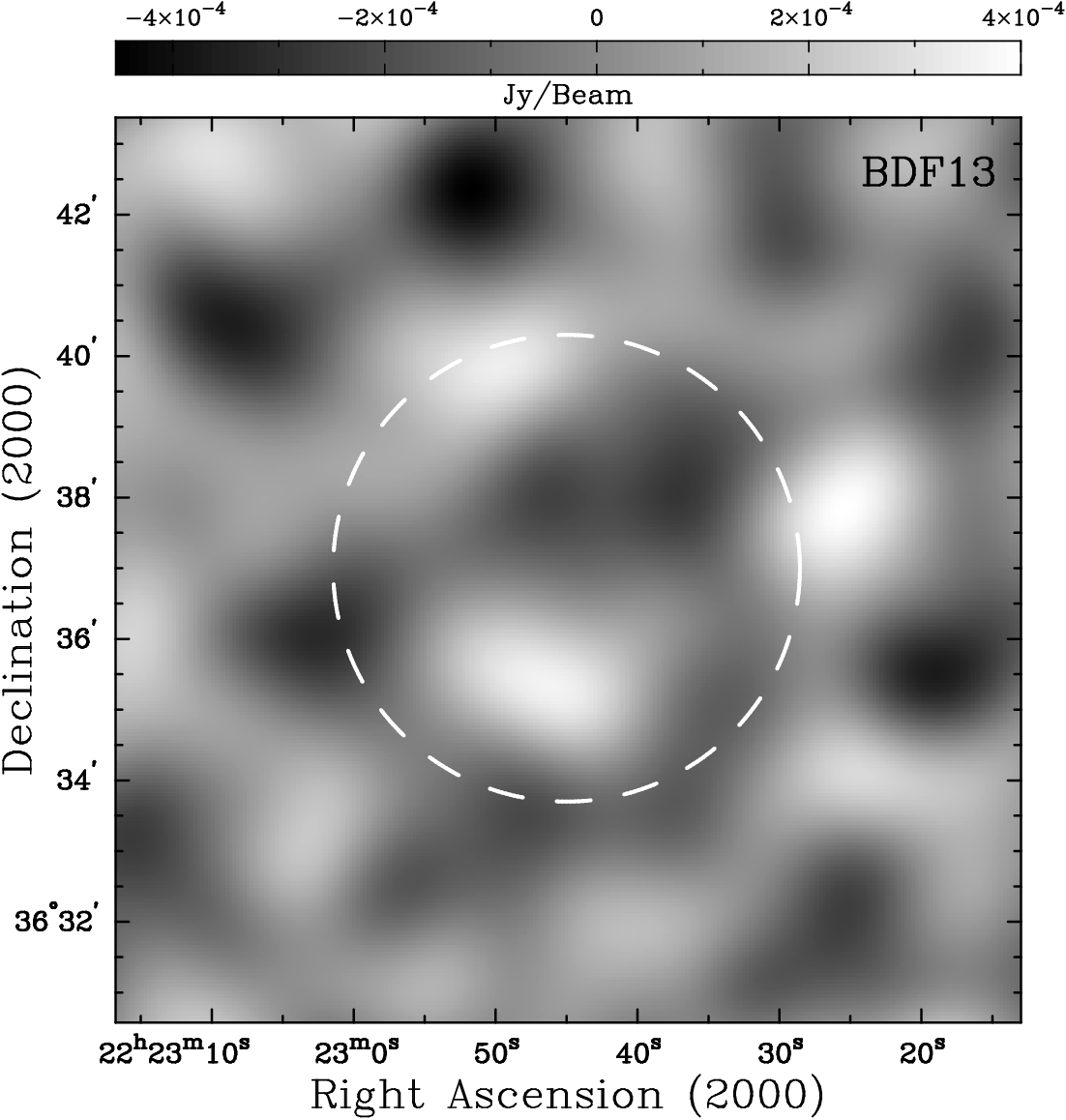}
  \includegraphics[scale=0.3]{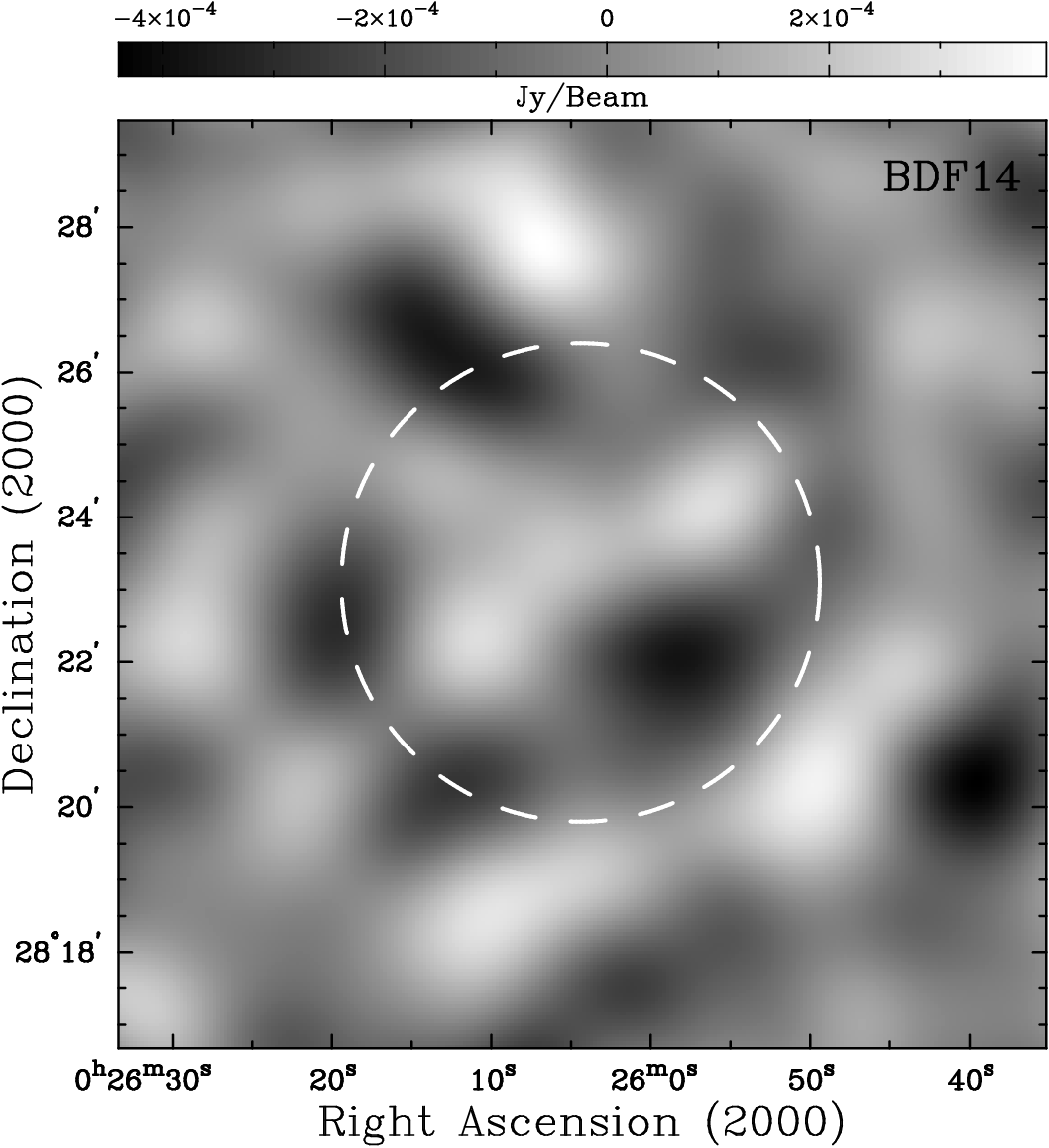}
  \includegraphics[scale=0.3]{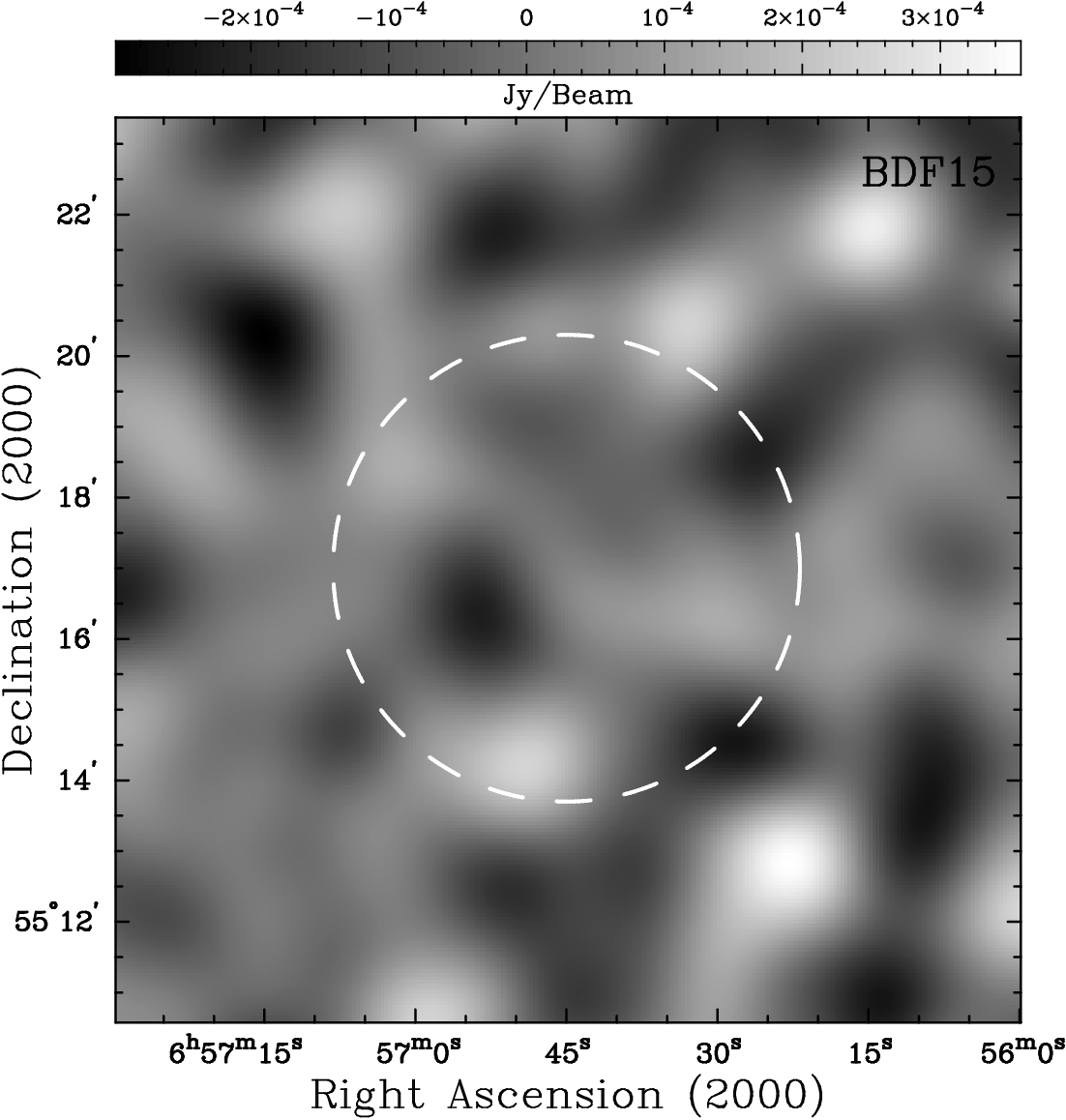}
}
\centerline{
  \includegraphics[scale=0.3]{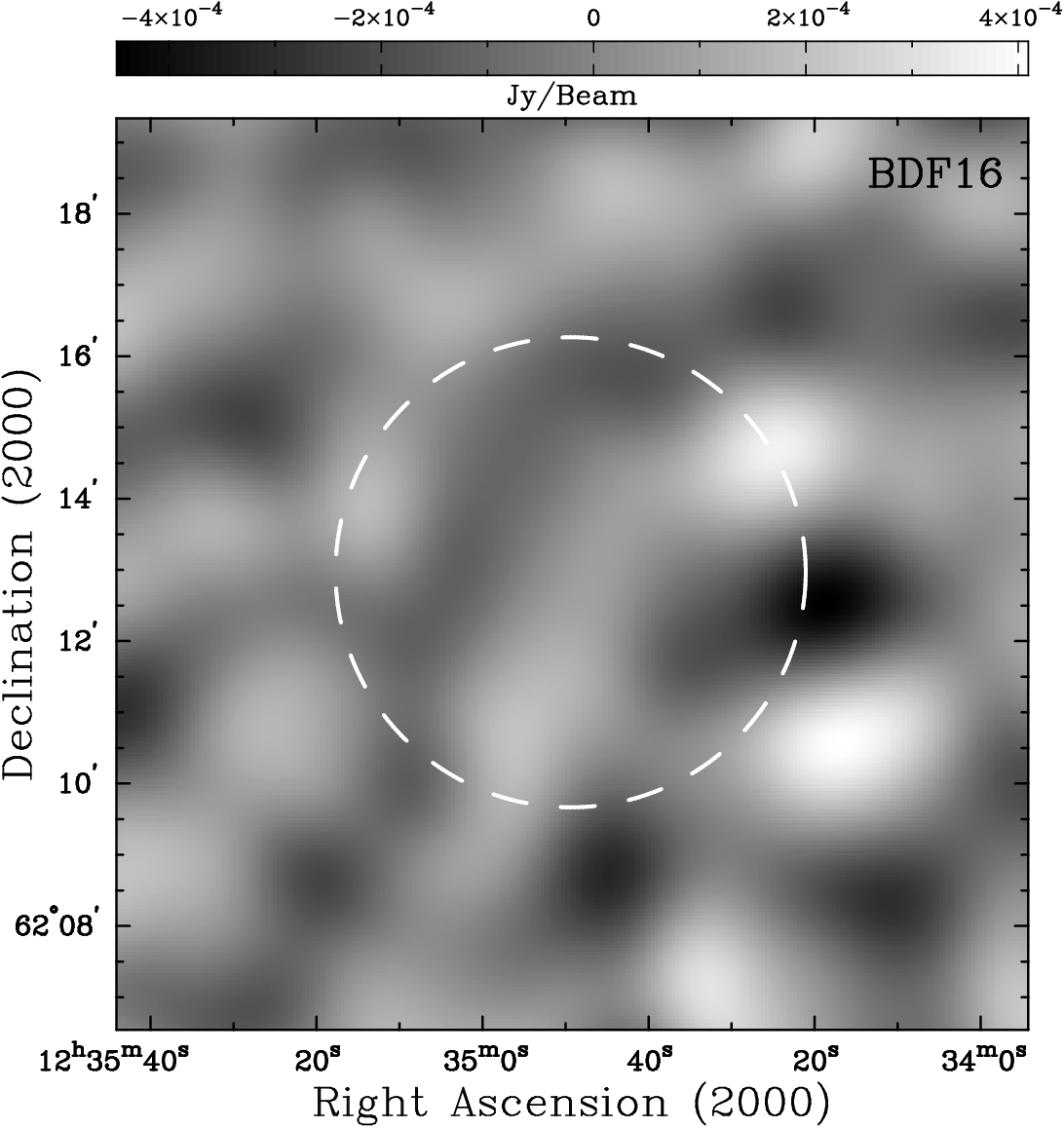}
  \includegraphics[scale=0.3]{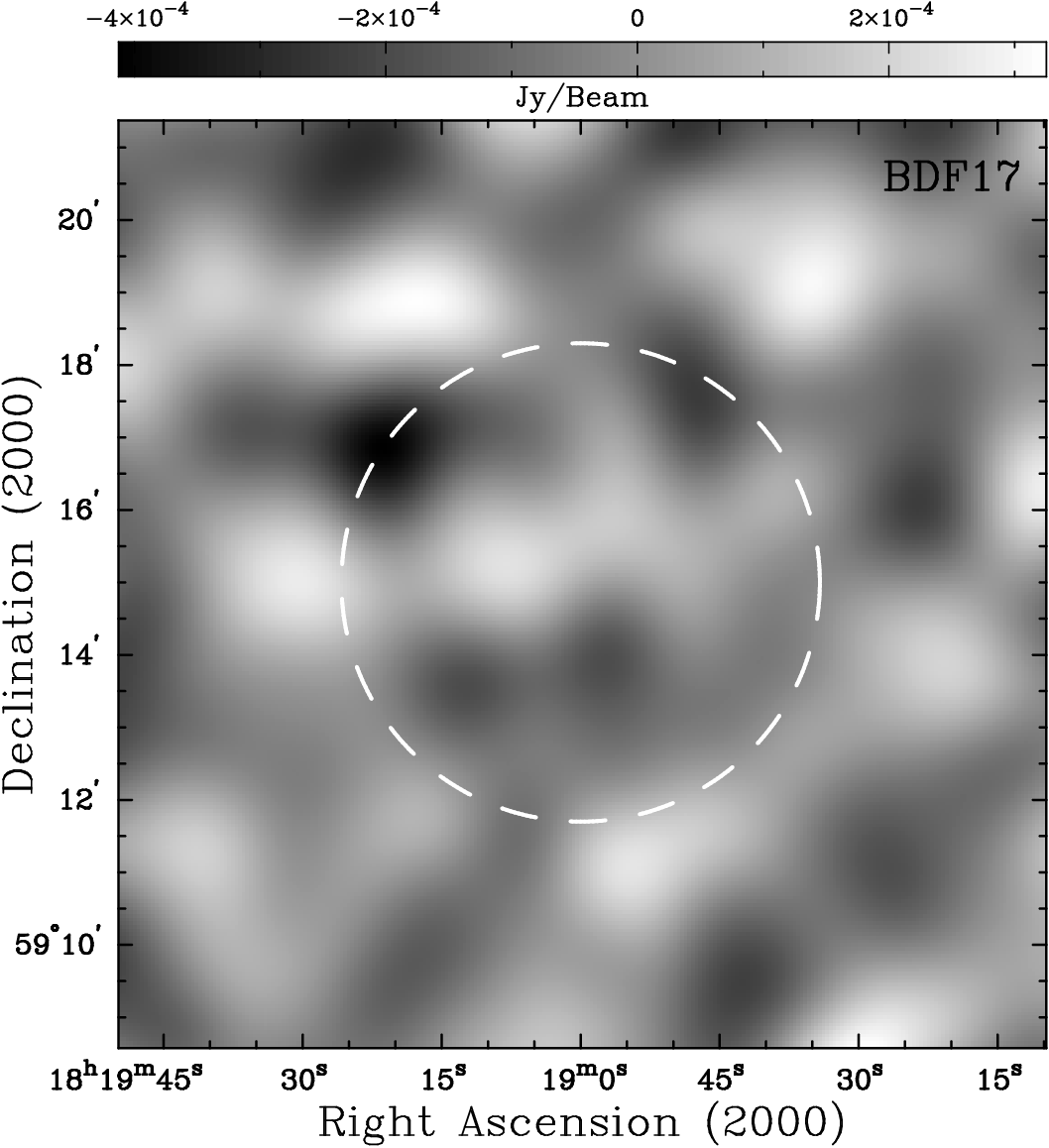}
  \includegraphics[scale=0.3]{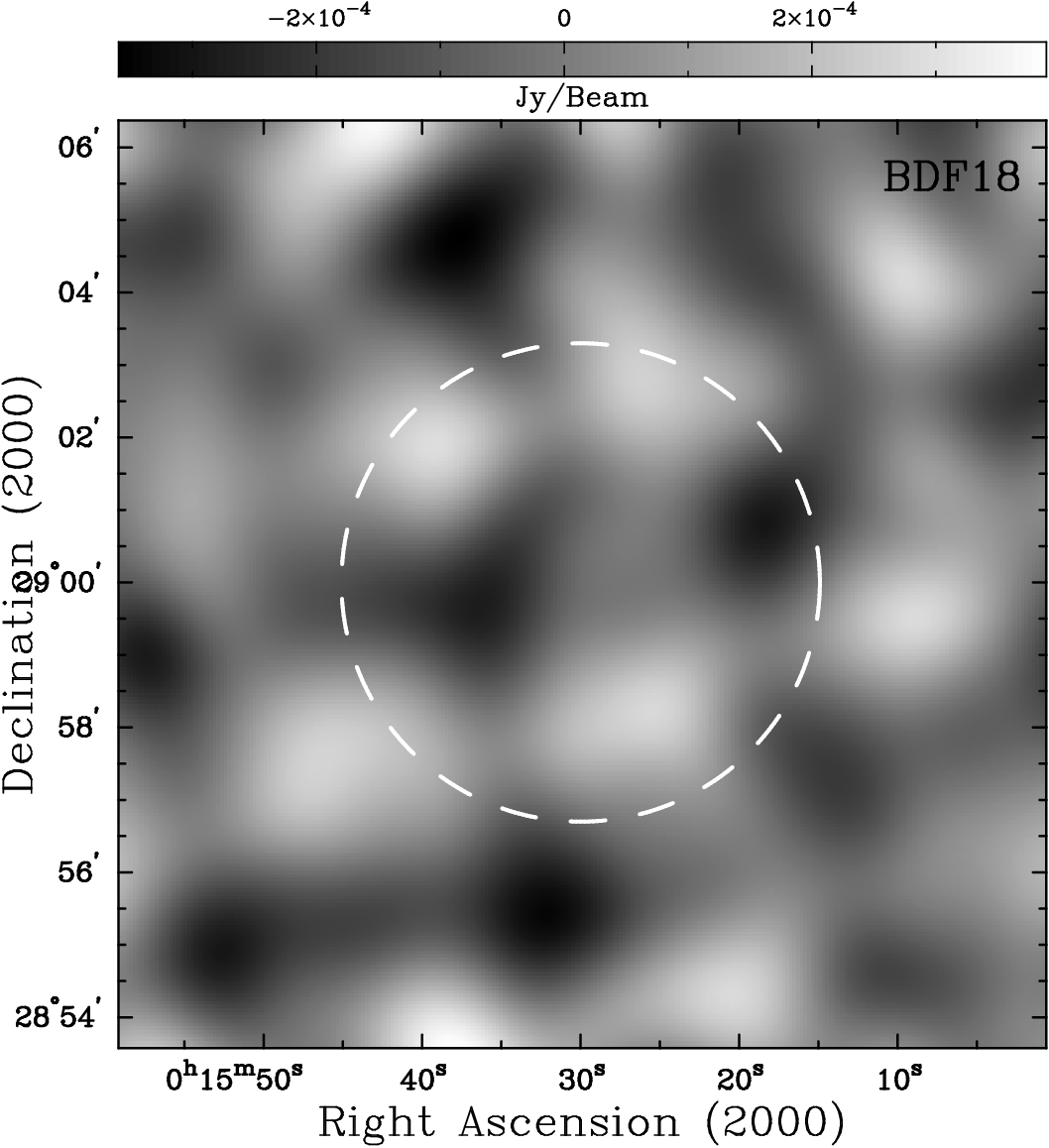}
}
\centerline{
  \includegraphics[scale=0.3]{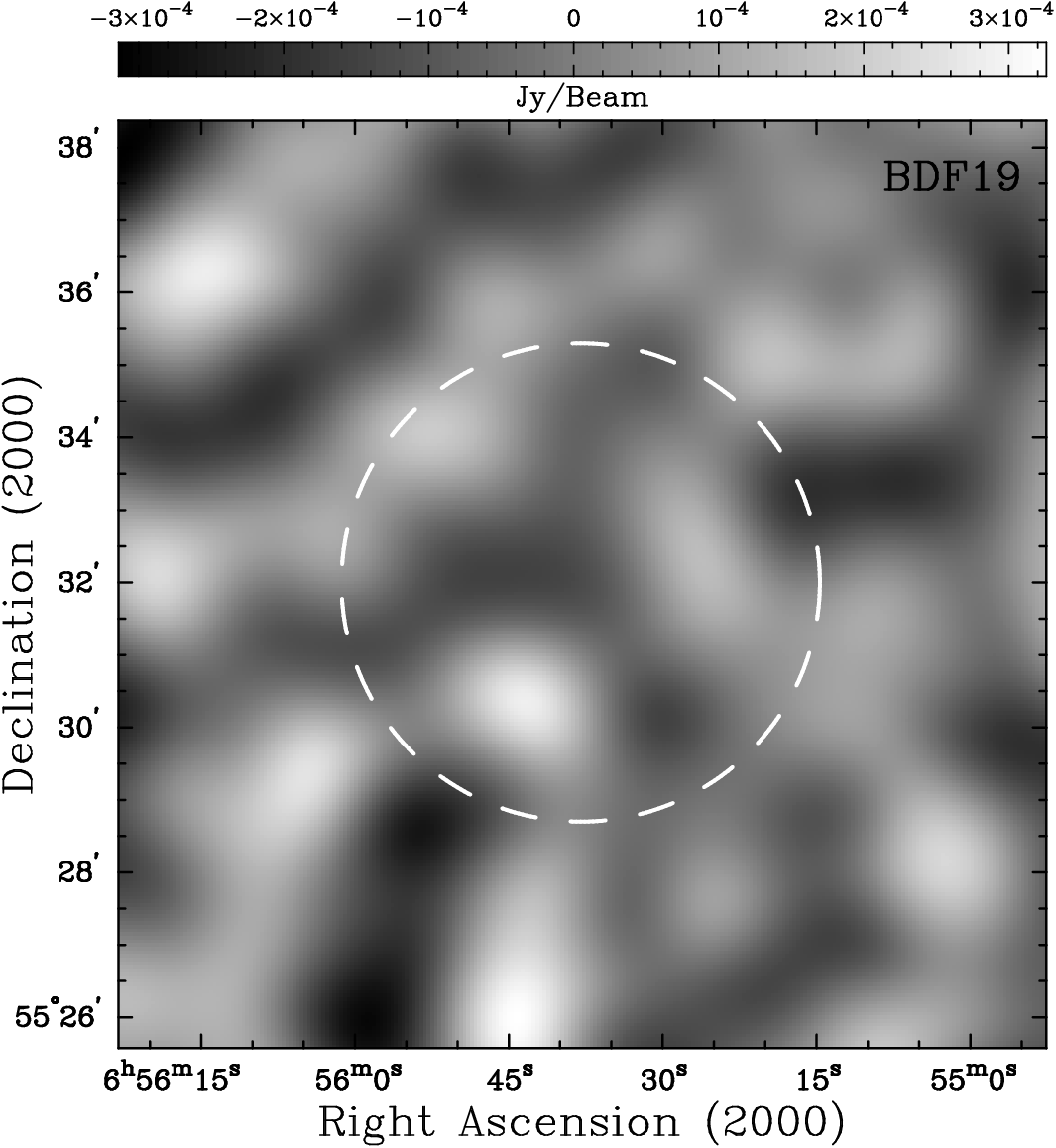}
  \includegraphics[scale=0.3]{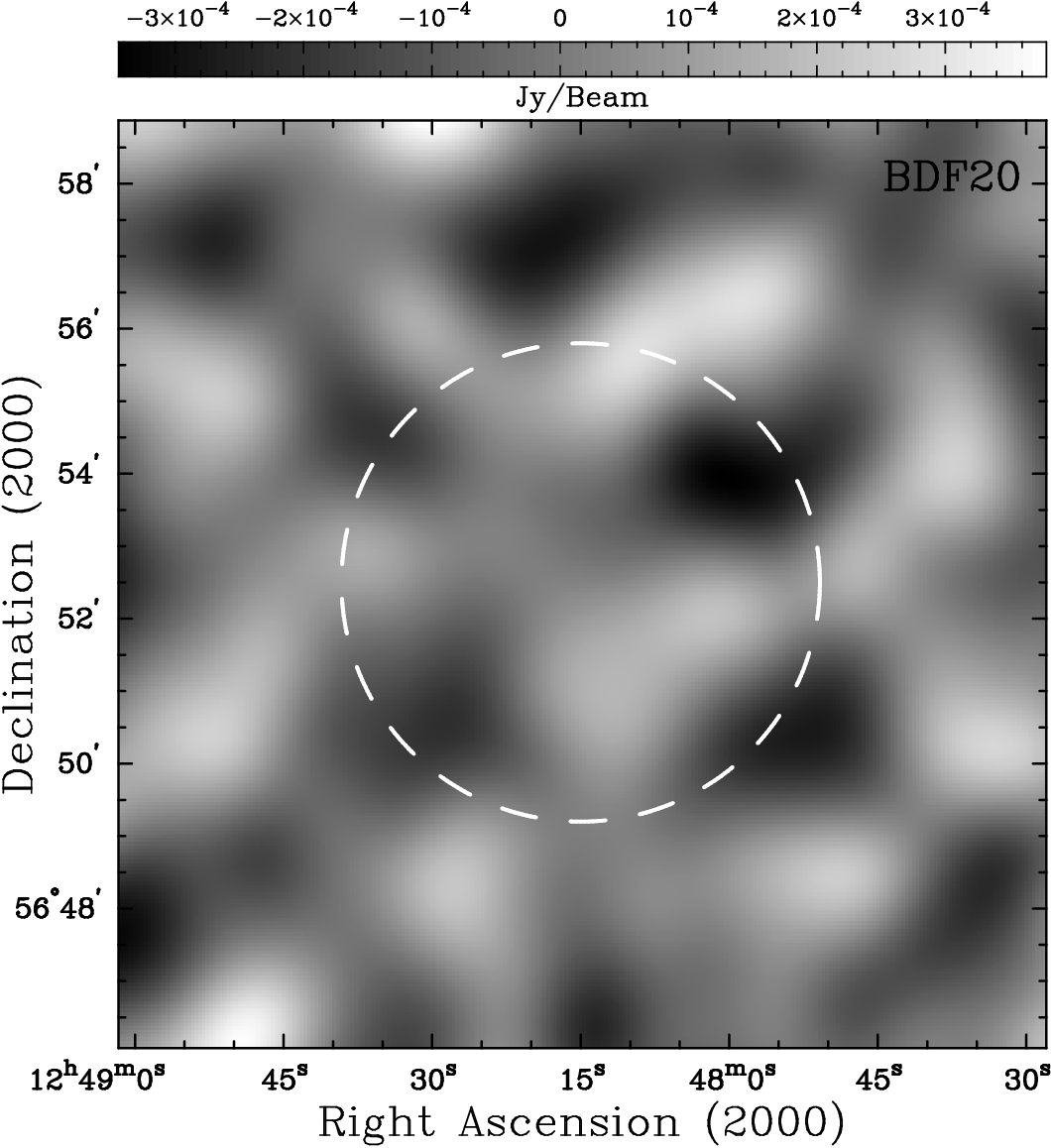}
  \includegraphics[scale=0.3]{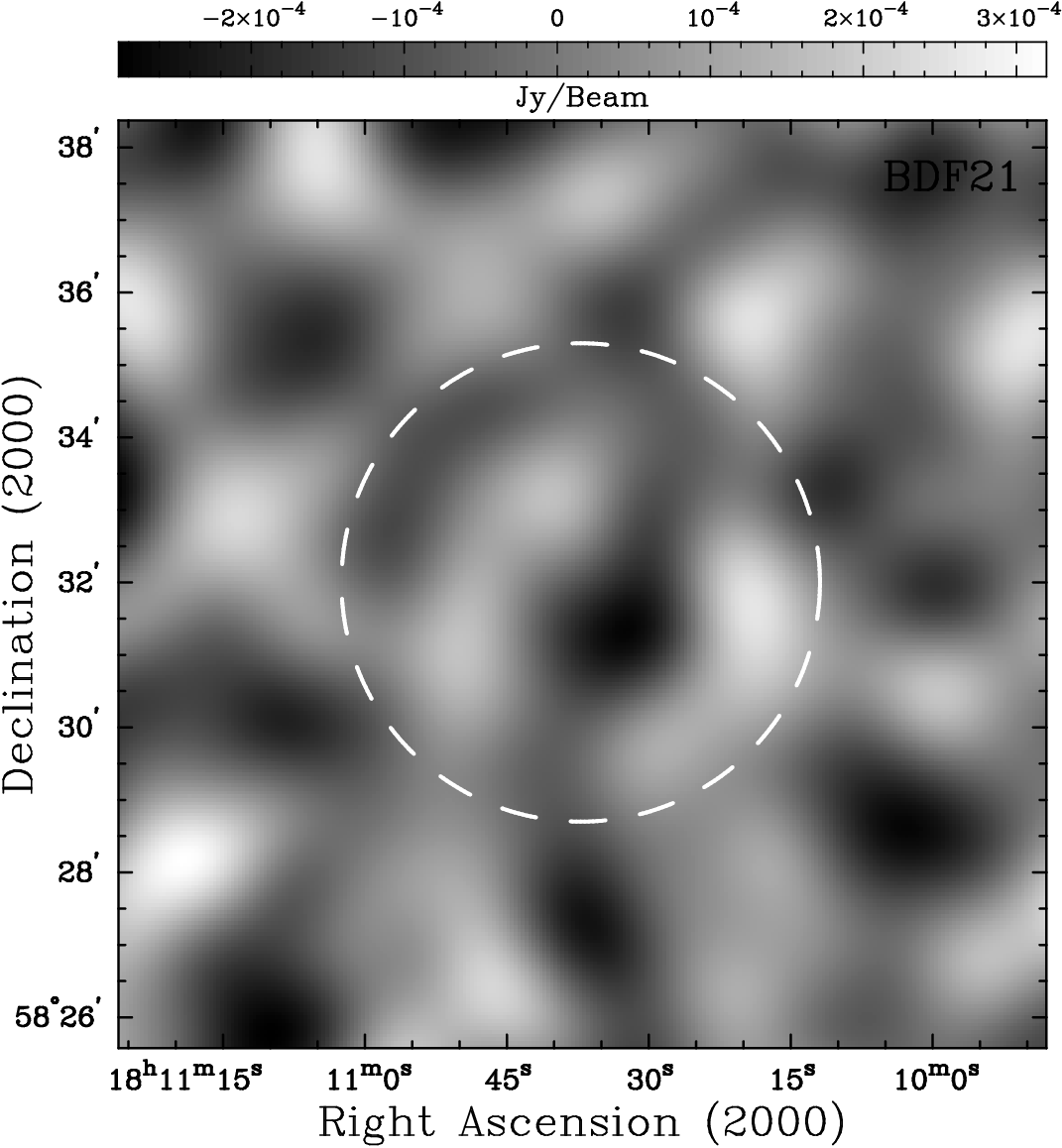}
}
\contcaption{Contd.
\\
}
\end{figure*}

\clearpage
 
\subsection{Image Simulations} \label{subsec:mc}

Decrements caused by galaxy clusters would be expected to produce an
excess of high SNR negative pixels.  We attempt to detect this unique 
signature using higher order statistics of the image flux distribution
as described in Rubino-Martin \& Sunyaev, 2003.
We compute asymmetry, skewness, and extrema in the observed data set.
These results are compared to both simulations of instrumental noise and to
simulations of SZ galaxy clusters added to instrumental noise in order to
determine the significance of the observed distribution of flux.
 
The Fourier transform of the visibilities measured by the
interferometer into the image plane introduces 
correlations between pixels which complicate the noise properties.  In
order to better understand this, we generated Monte Carlo
simulations of the visibility data and transformed them into images.
We created two sets of simulations, each containing 100 realizations of the full set of 18 images.
These simulations are used
to quantify the significance of applying the statistical tests to
the BIMA images.  The first set of
simulations uses random complex visibilities with variances consistent
with the observed noise for each image.  At each point in $u$-$v$ space
in the observed data, a simulated visibility is created with a
real and imaginary component from a random sampling of a Gaussian
distribution with variance determined from the weight
of that visibility.  Therefore, each simulated observation has $u$-$v$ coverage
and noise characteristics identical to the real BIMA observation of
each field.
 
In the second case, we add the expected contribution from the SZ sky to the MC noise realizations
using the same simulated SZ sky images described in Section~\ref{subsec:sig8} with $\sigma_8$ fixed to $1.0$. 
Independent regions of the
N-body simulations are attenuated with the $6.6\pr$ FWHM primary beam and
transformed into the $u$-$v$ plane using the exact same $u$-$v$ sampling as that
in the real BIMA data.  Noise is then added to the visibility data in
the same way as for the noise-only simulations and
the data is then transformed into the image plane for analysis.
 
The image statistics of the observed images are compared with those generated
from the Monte Carlo simulations.
For each simulation, an image is
generated from the raw visibilities tapered in the $u$-$v$ plane with
a $1.0 \, \kl$ FWHM Gaussian.  Statistics are generated directly from
the images of the 100 simulations.  Using a large number of Monte
Carlo simulations, one can produce a probability distribution for each
observable statistic.  The most likely value and integrated confidence
intervals can be taken from this distribution and compared to the results 
of the real observations.
Due to the modest number of 100 simulations of the
survey, the shape of the pixel distribution is strongly
dependent on bin size and bin spacing, leaving the most likely value
as a poorly determined quantity.  Instead, the median value in the 100
simulations is taken to be the most likely estimate.  The terms
defining the $16$th and $84$th percentile in the distribution are
taken to be the lower and upper bounds to the $68\%$ confidence
intervals, respectively.  The 4 nearest neighbors to the median, lower
bound, and upper bound in the distribution are averaged to
reduce the noise in the estimates.
 
\begin{figure*}[!h]
\centerline{
  \includegraphics[scale=0.45]{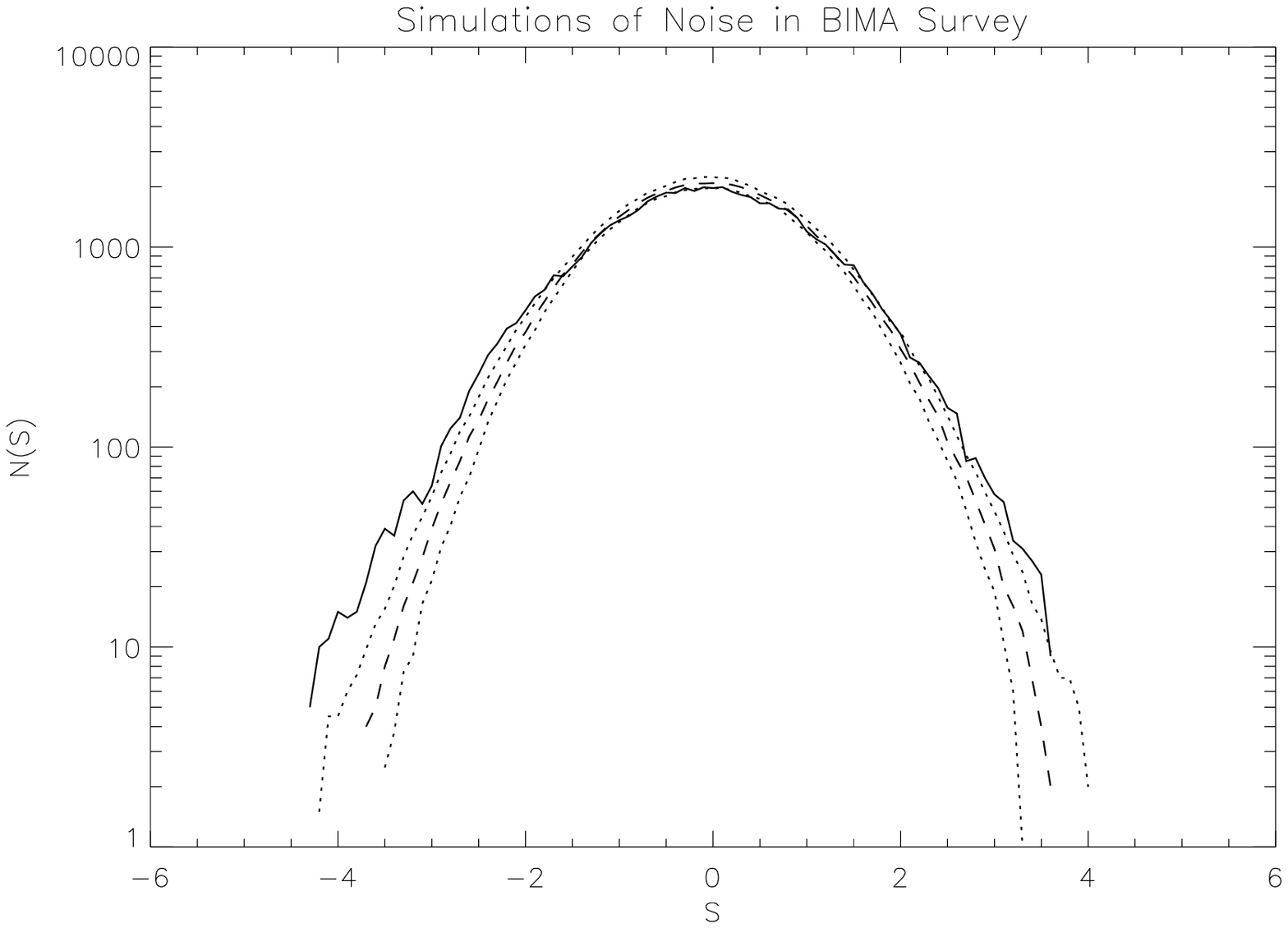}
  \includegraphics[scale=0.45]{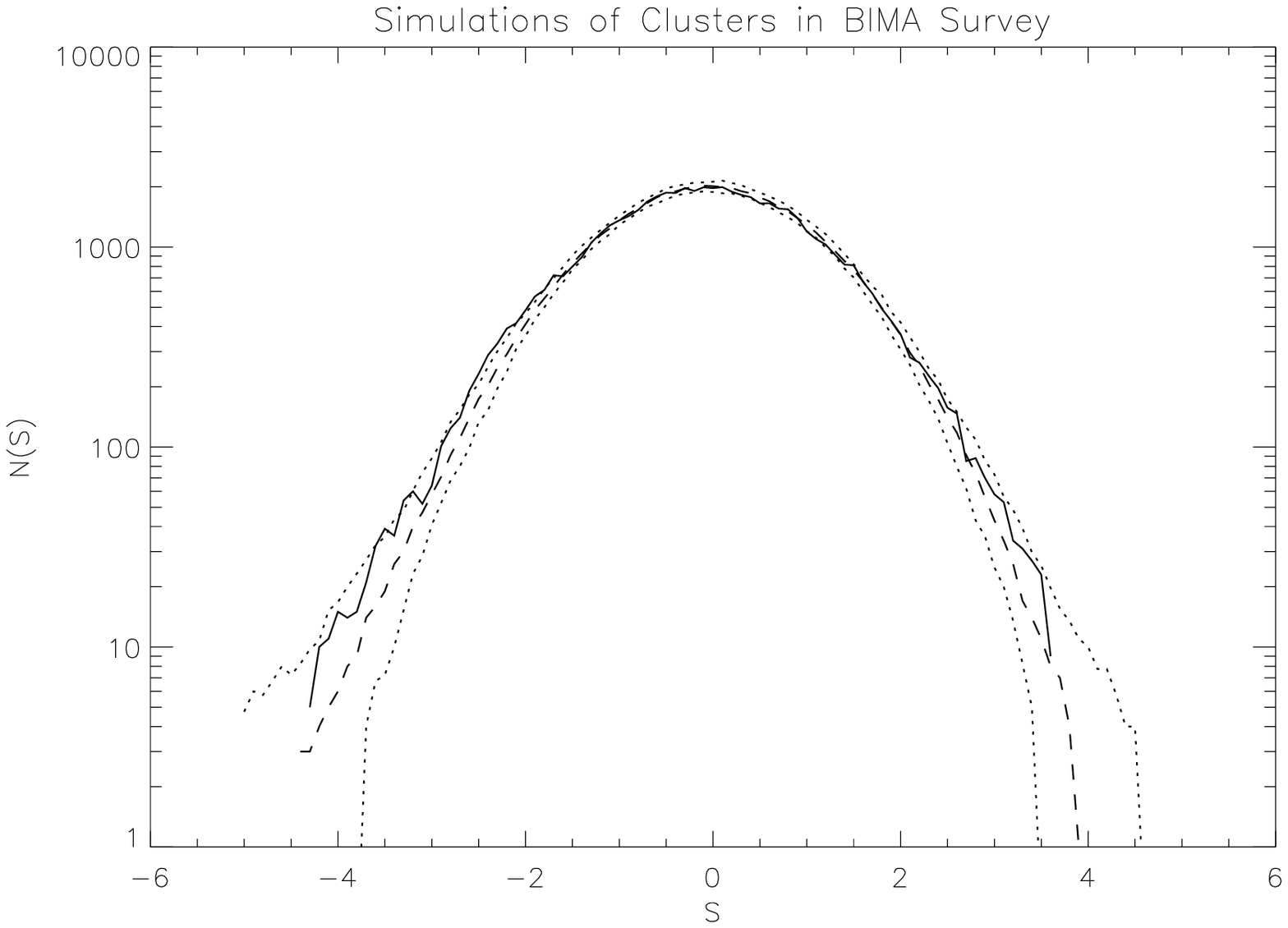}
}
%\plottwo{imstat/f26.ps}{imstat/f27.ps}
\caption[Distribution of pixels in simulations of BIMA survey]{
\\
The figure on the left represents the histogram for simulations including only instrumental noise.  
The figure
on the right represents the simulations with clusters added.  In both cases, the
solid line represents observed data, the dashed line represents the most likely value from the simulations,
and the dotted lines represent the $68\%$ confidence interval of the simulations.
The simulations including clusters are a better fit to the observed data not only in 
the region of negative pixels,
where the SZE is expected to directly contribute to the signal,
but also in the region of positive pixels where the effects of the negative 
sidelobes contribute to the signal.
}
\label{fig:Distribution_survey}
\end{figure*}

We use a histogram of pixels binned in
intervals of significance, $S=x_i/\sigma$, where $x_i$ is the value at pixel i and $\sigma$
is the estimated image RMS (approximately $110 \, \mu$Jy/beam).
This is equivalent to weighting the pixels by the
sensitivity of the observation.  
We can compare the pixel distributions for observed data and simulations
by comparing the histogram of the images as
shown in Figure \ref{fig:Distribution_survey}.

\clearpage
 
\subsection{Asymmetry of Pixel Distribution}\label{subsec:ass}
 
\begin{figure}[!htb]
  \plotone{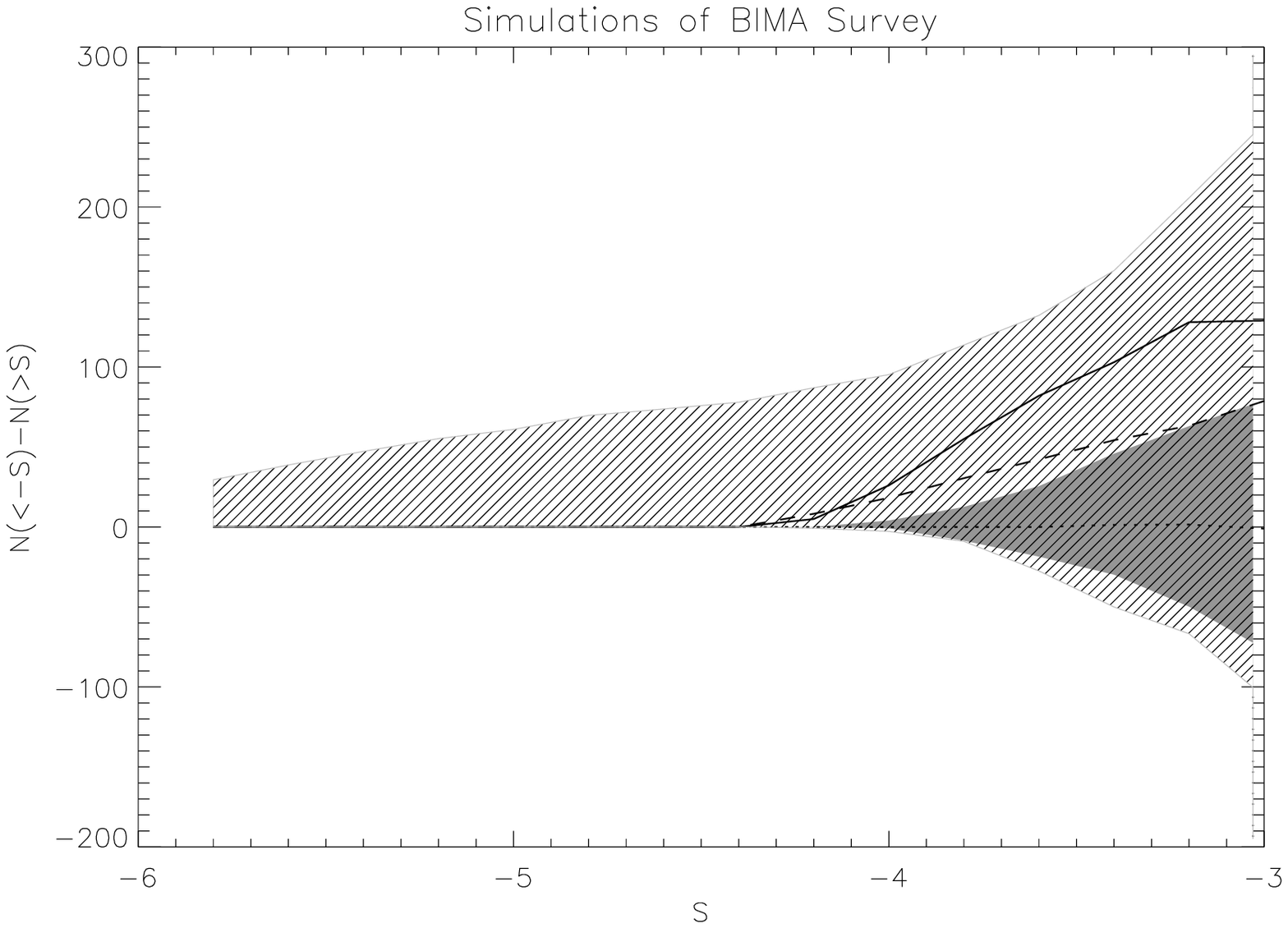}
\caption[Asymmetry in pixel distribution in simulations of BIMA survey]{
\\
This figure describes the asymmetry of the pixel distribution as a function
of SNR. The dark shaded region and dotted line represent the $68\%$ confidence intervals and median
asymmetry of the simulations that include only instrumental noise.
The region of angled parallel lines and the dashed line represent the $68\%$ confidence intervals and median
asymmetry of simulations that also include the SZE.
The black solid line represents the observed data.
}
\label{fig:Asym_survey}
\end{figure}

For a given image, we can characterize the pixel flux distribution
by selecting a flux interval $\Delta D$, and computing a
histogram of the number of pixels with a flux between $D-\Delta D/2$ and
$D+\Delta D/2$).  The asymmetry of this histogram can
be estimated directly as the difference in area between the positive
and negative regions.  A comparison of this statistic for the
BIMA images and simulations is shown in Figure \ref{fig:Asym_survey}.
 
It should be noted that the cluster simulations produce an excess of
positive pixels compared to simulations of instrumental noise only.
There are no sources with positive flux included in the
cluster simulations, however pixels with positive flux are expected
due to the sidelobes of the synthesized beam.
A Gaussian taper was applied to the $u$-$v$ data to minimize the sidelobes and
this effect.
 
It is clear from Figure \ref{fig:Asym_survey} that the observed data is 
consistent with the asymmetry in the simulations that include galaxy
clusters and inconsistent with simulations that include only noise.
For example, the excess of pixels with $S<-3$ in the observed data
exceeds that in the simulations of instrumental noise at $86\%$
confidence.
The number of pixels with $S<-4$ exceeds that found in the simulations of only instrumental noise at $88\%$ confidence.  
Therefore, the application of this statistic results in a detection of 
signal with the morphology of the SZ effect
with a significance 
comparable to the significance of the
detection of excess power reported in Section~\ref{sec:power}.
In fact, the observed asymmetry in the BIMA data exceeds the mean value from the noise \& SZ 
simulations.

As a simple exercise, a similar analysis is used to rule out positive
flux (such as point sources)
as the cause of excess power with slightly less confidence;
we simply repeat the 
analysis described, but with the sign of the flux from clusters reversed.
The excess of pixels with $S<-3$ in the observed data
exceeds that in the simulations using positive flux at $74\%$
confidence.
The excess of pixels with $S<-4$ exceeds that in simulations of instrumental noise at $76\%$ confidence.
A more rigorous analysis for point sources 
would include a more realistic distribution
of point sources and take into account the details of their removal using the full, 
equally weighted, set of visibilities.
This expanded analysis is considered unnecessary given the results of the VLA observations,
constraints on point sources, and lack of power at finer angular scales.
Nonetheless, this method would be useful in cases where rejection of point sources was
less certain.
We again conclude that it is very unlikely that point sources are responsible for the asymmetry 
observed in the BIMA image pixel fluxes.

\subsection{Skewness of Pixel Distribution}\label{subsec:skew}
 
\begin{figure}[!htb]{}
  \plotone{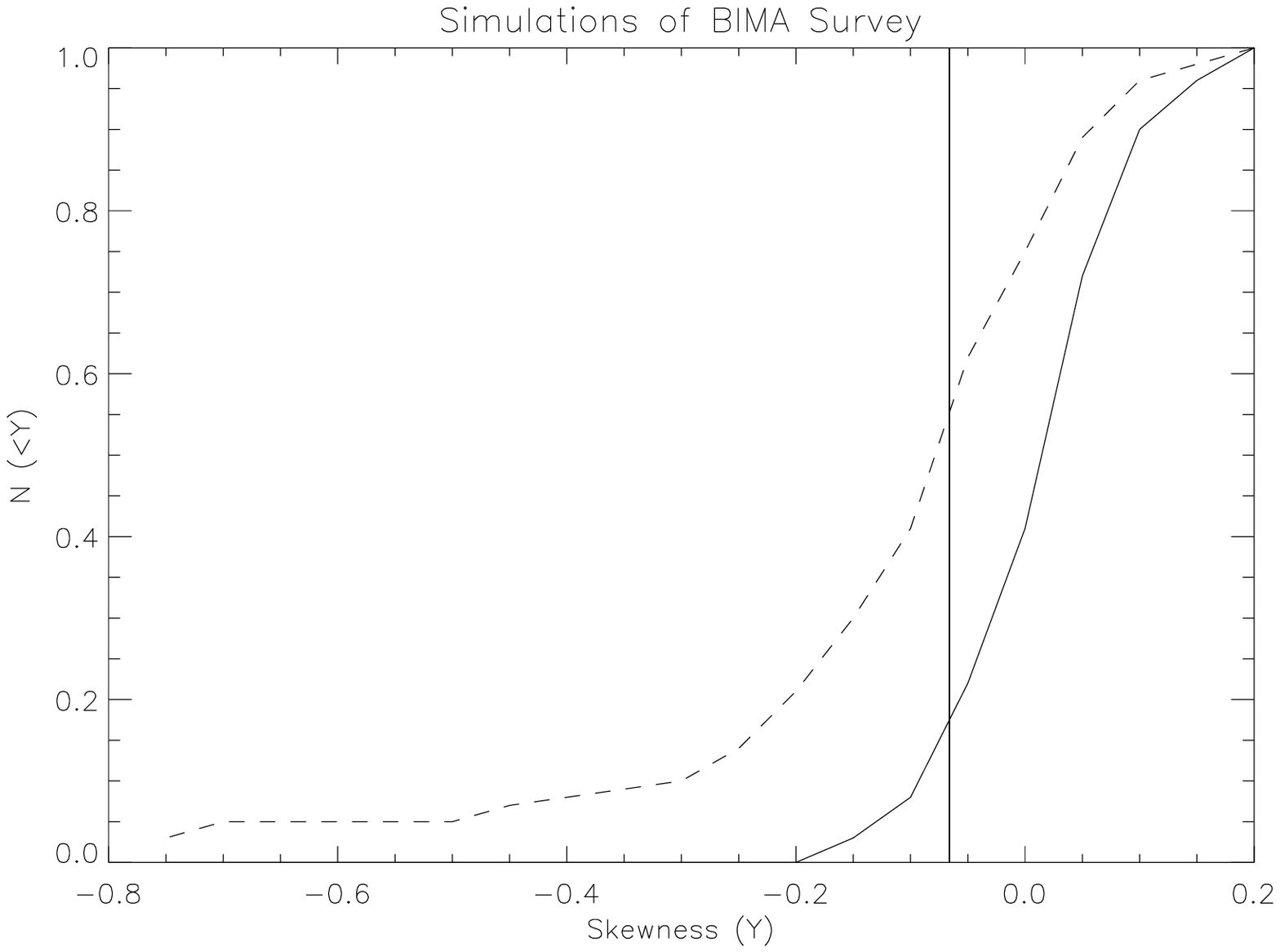}
\caption[Skewness of pixel distribution in simulations of BIMA survey]{
\\
The cumulative count of simulations as a function of skewness.  The
solid line represents the model assuming only instrumental noise and the dashed line 
represents the model including the simulated SZ sky from the n-body simulations.
The vertical line represents the skewness measured in the BIMA data.
}
\label{fig:Skewness_survey}
\end{figure}

A measure of skewness conveys
information about the sign of the features producing
the deviation from Gaussianity. This quantity can be estimated
using the third moment of the data:
\begin{equation}
Y = \frac{1}{N_{pix}} \sum_{i=1}^{N_{pix}} (x_i - \overline{x})^3
\label{skew}
\end{equation}
The skewness of the data derived from the BIMA images is compared to the skewness determined from the
Monte Carlo simulations in Figure \ref{fig:Skewness_survey}.
 
Analysis of the simulations of instrumental noise alone results in a skewness of 
$0.01\pm0.08$ at $68\%$ confidence, while the simulations of clusters plus instrumental noise
give a mean skewness of $-0.096^{+0.11}_{-0.15}$.
The skewness in the BIMA observations was found to be $Y = -0.066$, a value that is 
inconsistent with the noise only simulations at $64\%$ confidence. 
Repeating the simple analysis described at the end of \S\ref{subsec:ass},
the skewness is inconsistent with positive flux sources as the cause of the observed 
excess power at $84\%$ confidence.

\subsection{Outliers in the Distribution}\label{subsec:out}
In addition to the two tests described above, we performed simulations to characterize 
the significance of outliers in pixel flux distributions of the individual fields.  
We identified the most negative pixels in each of the 18 BIMA fields and compared
these with the results of the simulations.
Seven BIMA fields were observed to have decrements with $S <-3\sigma$, more than occurred
in $93\%$ of the survey simulations with noise only.
Two BIMA fields were observed to have decrements with $S<-4\sigma$, more than occurred
$99\%$ of the survey simulations including only noise.
None of the observed BIMA fields had a pixel with positive flux $S >4\sigma$.

\subsection{Interpretation of Image Statistics}\label{subsec:im_sum}
In each of the statistical tests described above, the BIMA images were
found to be inconsistent with those produced by instrumental noise
alone at approximately the $1-2 \sigma$ level, comparable to the
significance of the detection of excess power.  Although the analysis
does not conclusively determine the source of the excess power, the
results are consistent with the signal expected from SZ galaxy
clusters. 
The results are also inconsistent with radio point sources being the source
of the observed excess power. 
Despite the lack of a definitive conclusion from these tests, the results are
encouraging.
In order to investigate what observations would be required to make a definitive measurement,  
we performed simulations of the 18 field survey with four times the observation time,
or equivalently four times the correlation bandwidth.
These simulations resulted in a detection of excess power and skewness
at the $4 \sigma$ level in more than $50\%$ of the simulations.  Such
a detection would provide convincing evidence for SZ clusters as the
source of excess power.  

\section{Optical Observations of BIMA Fields}\label{sec:optical}

To date, all known galaxy clusters have been discovered through 
optical or X-ray observations.  A negative correlation of the observed 
fine scale CMB anisotropy with X-ray or optical emission would be a smoking gun for
the discovery of a cluster through the SZ effect.  To search for this
correlation, we observed ten of the BIMA fields using ground-based optical telescopes.
We selected the five fields with significant levels of excess power
assuming that they are the most likely candidates for identifying
galaxy clusters.  In addition to those five fields, we observed five
BIMA fields that lie at convenient RA during the nights we were
awarded time.  Imaging was performed using I and R filters on the LRIS
instrument (Oke \ea 1995) on the 10 m Keck\footnote{Keck
Observatory is operated as a scientific partnership among the
California Institute of Technology, the University of California and
the National Aeronautics and Space Administration.  The Observatory
was made possible by the generous financial support of the W.M. Keck
Foundation.} I telescope.  Imaging in z' was done with the MOSAIC
instrument (Wolfe \ea 1998) on the 4m Kitt Peak National
Observatory telescope\footnote{KPNO is a Division of the National
Optical Astronomy Observatory, which is operated by the Association of Universities for Research in Astronomy,
Inc. (AURA) under cooperative agreement with the National Science Foundation.}.

We performed an analysis of the optical images using a method similar to that
used in the Red Cluster Survey
(RCS) survey (Gladders \& Yee, 2000).
We first categorized objects into two redshift bins determined by color
$R-I$ and $R-z'$.
The first bin contains only high redshift galaxy candidates.
These candidates have color $R-I > 1.0$ and
$R-z' > 1.0$, implying a red sequence redshift $0.5 < z <1.0$.
The second bin contains low redshift galaxy candidates.
These candidates have color $0.0 < R-I < 1.0$ and
$0.0 < R-z' < 1.0$ implying a redshift $z < 0.5$.
All objects not satisfying these criteria are considered foreground contamination and are
discarded.
An exponential kernel is used to smooth the maps of detected objects in each redshift bin with scale radius of $20\twopr$
to create a surface density map.
We then create a product map by simply multiplying each surface density map with the corresponding BIMA map
described in \S\ref{sec:images}.
The product maps are searched for peaks and asymmetry in the distribution.
Statistics are quantified with Monte Carlo simulations of random BIMA fields.
The resulting analysis (described in full detail in Dawson, 2004)
showed no significant correlation between overdensities of galaxies in the optical
maps and decrements of flux in the BIMA $28.5$ GHz maps.

While the follow-up optical observations do not confirm the
hypothesis that the observed anisotropy is caused by the SZ effect in
galaxy clusters, 
the method of combining optical and SZ observations should prove to be a powerful 
technique for identifying galaxy clusters in future SZ surveys.  
In this first attempt at doing so, we have not modeled the expected optical
cluster signature sufficiently to say what constraints this null
result places on the role of SZ clusters in producing the observed
excess power.
There also remains the possibility that anisotropy from galaxy clusters at
redshifts beyond the sensitivity of the optical data,  $z>1$, could be contributing significantly
to the observed signal in the BIMA survey.

\section{Conclusion}\label{sec:con}
In this paper, we report the final results from our search for arcminute scale
CMB anisotropy using the BIMA array.
Modeling the observed power spectrum with a single flat band power with average
multipole of $\ell_{eff} = 6864$, we find
$\Delta T^2=170^{+120}_{-100}\,\mu$K$^2$ at $68\%$ confidence and a detection of
$\Delta T^2 >0$ at $92.3\%$ confidence.
Dividing the data into two bins corresponding to different spatial resolutions
in the power spectrum, we find
$\Delta T_1^2=220^{+140}_{-120}\,\mu$K$^2$ at $68\%$ confidence for CMB flat band power
described by an average multipole of $\ell_{eff} = 5237$ and
$\Delta T_2^2<840\,\mu$K$^2$ at
$95\%$ confidence for $\ell_{eff} = 8748$.
We have used VLA observations and various cuts to test for contamination from
radio point sources and systematic effects
and conclude that it is unlikely that these sources are responsible for the observed signal.
If we assume that the measured excess power is due to a background of distant SZ clusters,
we can compare its value with that from simulations of large scale structure to place a constraint 
on the normalization of matter fluctuations, $\sigma_8=1.03^{+0.20}_{-0.29}$ at $68\%$
confidence. 

In order to try to determine the source of the observed anisotropy power, we have performed
an analysis of the BIMA image statistics.
We compared the skewness, asymmetry, and outliers of the measured pixel flux distribution with
simulations including noise only and noise plus SZ clusters.
A statistical analysis of the BIMA survey images found that they were consistent with simulations
including a background of SZ clusters, and inconsistent with simulations of instrumental noise alone 
or noise plus radio point sources at $1-2\sigma$.
Additional Monte Carlo simulations indicate that with approximately four times the time dedicated 
to the survey, or equivalently four times the correlated bandwidth,
the BIMA instrument would achieve the sensitivity to test the hypothesis of SZ clusters
as the source of the observed excess power at greater than $99\%$ confidence.
Therefore, future dedicated interferometers, such as the Sunyaev-Zel'dovich 
Array\footnote{http://astro.uchicago.edu/sza/}, should be able to effectively use 
image statistics to determine if any observed anisotropy is due to SZ clusters.
Finally, we performed a preliminary search for a correlation between red galaxy density 
and CMB temperature fluctuations.
We are currently unable to quantify the significance of the null result, however, 
we expect that X-ray and optical follow-up will be essential tools for the interpretation
of future SZ surveys. 

\acknowledgments

We thank the entire staff of the BIMA observatory for their
many contributions to this project, in particular
Rick Forster and Dick Plambeck for their assistance with both the
instrumentation and observations.
Brian Wilhite, Josh Simon, and Steve Dawson are thanked for their early advice on optical
observations and data reduction.  We would also like to thank Mike Gladders for his suggestions regarding
identification of clusters in the optical data
and Ramon Miquel, Radek Stompor, and Chao-lin Kuo for their stimulating discussions of Bayesian statistics.
We are grateful for the scheduling of time at the VLA, Keck, and KPNO observatories 
in support of this project.
This work was supported in part by NASA LTSA grant number NAG5-7986,
NSF grants AST-0096913 and PHY-0114422, and the David and Lucile Packard Foundation. 
The BIMA millimeter array is supported by NSF grant AST 96-13998.

Facilities: \facility{BIMA}, \facility{VLA}, \facility{Keck}, \facility{KPNO}.

\end{document}